\documentclass[aps, prd, twocolumn, nofootinbib,]{revtex4-2}

\usepackage{amsmath}
\usepackage{amssymb}
\usepackage{amsfonts}
\usepackage{wasysym}
\usepackage{graphicx}
\usepackage{comment}
\usepackage{tensor}
\usepackage{cancel}
\usepackage[svgnames]{xcolor}

\usepackage[colorlinks = true, linkcolor = DarkBlue, citecolor = DarkBlue, urlcolor = DarkBlue, bookmarks = false]{hyperref}

\def\filetype{pdf}
\def\path{}

\allowdisplaybreaks[1]

\begin{document}


\title{Einstein-Dirac system in semiclassical gravity}
\author{Ben Kain}
\affiliation{Department of Physics, College of the Holy Cross, Worcester, Massachusetts 01610, USA}

\begin{abstract}
\noindent
We study the Dirac equation minimally coupled to general relativity using quantum field theory and the semiclassical gravity approximation.  Previous studies of the Einstein-Dirac system did not quantize the Dirac field and required multiple independent Dirac fields to preserve spherical symmetry.  We canonically quantize a single Dirac field in a static spherically symmetric curved spacetime background.  Using the semiclassical gravity approximation, in which the Einstein field equations are sourced by the expectation value of the stress-energy-momentum tensor, we derive a system of equations whose solutions describe static spherically symmetric self-gravitating configurations of identical quantum spin-1/2 particles.  We self-consistently solve these equations and present example configurations.  Although limiting cases of our semiclassical system of equations reproduce the multifield system of equations found in the literature, our system of equations is derived from the excitations of a single quantum field.
\end{abstract} 

\maketitle


\section{Introduction}

In the Einstein-Dirac system, the Dirac equation is minimally coupled to general relativity.  Solutions describe self-gravitating configurations of spin-1/2 particles.  This system was first studied in static spherically symmetric spacetimes by Finster et al.~\cite{Finster:1998ws}.  Additional studies of this system, including generalizations that couple the Dirac field to the electromagnetic field or to an $SU(2)$ gauge field, were undertaken in \cite{Finster:1998ux, Finster:1998ak, Finster:1998ju, Finster:2000ps, Nodari1, Nodari2, Stuart, Adanhounme:2012cm, Krechet:2014nda, Herdeiro:2017fhv, Dzhunushaliev:2018jhj, Dzhunushaliev:2019kiy, Blazquez-Salcedo:2019qrz, Dzhunushaliev:2019ham, Bronnikov:2019nqa, Blazquez-Salcedo:2019uqq, Dzhunushaliev:2019uft, Leith:2020jqw, BakuczCanario:2020qmq, Minamitsuji:2020hpl, Leith:2021urf, Leith:2022jew, Liang:2022mjo}.  Static and stationary axisymmetric spacetimes were considered in \cite{Bronnikov:2004uu, Herdeiro:2019mbz, Bronnikov:2009na} and dynamical solutions in time-dependent spherically symmetric spacetimes were found in \cite{Ventrella:2003fu, Zeller:2006rm, Daka:2019iix}.  Static spherically symmetric wormhole solutions in the Einstein-Dirac-Maxwell system were considered in \cite{Blazquez-Salcedo:2020czn, Konoplya:2021hsm,  Blazquez-Salcedo:2021udn}.

In all of these studies, an independent Dirac field is introduced in the Lagrangian for each particle that is in a distinct state.  Each field is then described by a classical solution to the equations of motion.  In some cases, motivated by the Pauli exclusion principle, the classical solutions are normalized, which implements a one-particle restriction for the respective field.  In this case, the classical solutions can be interpreted as first quantized wave functions.  In spherical symmetry, multiple Dirac fields are always introduced.  This is explained as being necessary to preserve spherical symmetry, in that only with multiple fields can the total angular momentum of the system be made to vanish.

In this paper, we introduce a \textit{single} Dirac field.  We canonically quantize the Dirac field using the formalism of quantum field theory in a curved spacetime background \cite{Birrell:1982ix, Wald:1995yp, Parker:2009uva}.  We preserve spherical symmetry by focusing on excitations of the vacuum with zero total angular momentum.  Using the semiclassical gravity approximation, in which the Einstein field equations are sourced by the expectation value of the stress-energy-momentum tensor \cite{Birrell:1982ix, Wald:1995yp, Parker:2009uva, Hu:2020luk}, we construct static spherically symmetric self-gravitating configurations of spin-1/2 particles in quantum field theory.  Our configurations are therefore populated by identical quantum particles.

The use of semiclassical gravity to describe self-gravitating spherically symmetric systems was recently undertaken by Alcubierre et al.~in a study of a single real scalar field \cite{Alcubierre:2022rgp}.  They found that excitations of the quantized real scalar field can be reduced to the classical Einstein-Klein-Gordon system.  In particular, the single quantum real scalar field can describe boson stars \cite{Kaup:1968zz, Ruffini:1969qy, Liebling:2012fv} as well as their relatives, such as $\ell$-boson stars \cite{Alcubierre:2018ahf} and configurations not previously considered in the literature.  Reference \cite{Alcubierre:2022rgp} was the inspiration for our study of the quantum Dirac field and, similar to \cite{Alcubierre:2022rgp}, we find that excitations of the quantized Dirac field can describe the configurations presented in \cite{Finster:1998ws}, as well as configurations not previously considered in the literature.

This paper is organized as follows.  In Sec.~\ref{sec:Einstein-Dirac}, we review the Dirac equation minimally coupled to general relativity.  In Sec.~\ref{sec:quantize}, we canonically quantize the Dirac field in a classical curved spacetime background and decompose the Dirac field operator in terms of mode functions, which are solutions to the classical equations of motion.  In Sec.~\ref{sec:eom sol}, we solve for the mode functions and show that they can be classified as being positive or negative frequency solutions.  This allows us to identify a preferred vacuum state and to construct spherically symmetric excitations of the vacuum state, which we do in Sec.~\ref{sec:states}.  In Sec.~\ref{sec:semiclassical gravity}, we briefly review the semiclassical gravity approximation.  We also present equations for determining the metric fields, which follow from the Einstein field equations, and compute the expectation value of the stress-energy-momentum tensor.  The complete system of equations, whose solutions describe static spherically symmetric semiclassical configurations of the Einstein-Dirac system, must be solved numerically.  In Sec.~\ref{ED SG}, we describe our numerical methods and present example solutions.  We conclude in Sec.\ref{sec:conclusion}.  Throughout, we use units such that $c=\hbar=1$.


\section{Einstein-Dirac}
\label{sec:Einstein-Dirac}

We study the static spherically symmetric Einstein-Dirac system.  For the static spherically symmetric spacetime, we parameterize the metric as
\begin{equation} \label{metric}
ds^2 = -\alpha^2(r) dt^2 + a^2(r) dr^2 + r^2 (d\theta^2 + \sin^2\theta d\phi^2).
\end{equation}
This metric describes a spacetime that is foliated by time-independent, and hence identical, spatial hypersurfaces $\Sigma$.  The induced spatial metric $\gamma_{ij}$ on $\Sigma$ and the future directed normal vector $n^\mu$ to $\Sigma$ are given by \cite{AlcubierreBook}
\begin{equation} \label{gamma n}
\gamma_{ij} = \text{diag} (a^2, r^2, r^2\sin^2\theta),
\qquad
n^\mu = (\alpha^{-1}, 0, 0, 0).
\end{equation}

In the matter sector, we have a single four-component Dirac spinor $\psi$ minimally coupled to gravity and described by the Lagrangian
\begin{equation} \label{Lagrangian}
\mathcal{L} = \sqrt{-\det(g_{\mu\nu})} \left( \bar{\psi} \gamma^\mu \nabla_\mu \psi - m_\psi \bar{\psi} \psi \right),
\end{equation}
where $\det(g_{\mu\nu})$ is the determinant of the metric, $\bar{\psi}$ is the adjoint spinor, $\gamma^\mu$ is a curved space $\gamma$-matrix, $m_\psi$ is the mass parameter, and $\nabla_\mu$ is a metric compatible covariant derivative.  

We couple the Dirac spinor to curved space using the vierbein formalism \cite{Weinberg:1972kfs, Carroll:2004st}, where the vierbein $e\indices{_a^\mu}$ is defined by
\begin{equation}
g_{\mu\nu} = e_{a\mu} e\indices{^a_\nu}, \qquad
\eta_{ab} = e_{a\mu} e\indices{_b^\mu},
\end{equation}
where $\eta_{ab} = \text{diag}(-1,1,1,1)$ is the flat space Minkowski metric.  Curved space $\gamma$-matrices, $\gamma^\mu$, are related to flat space $\gamma$-matrices, $\tilde{\gamma}^a$, through the vierbein,
\begin{equation} \label{gamma e gamma}
\gamma^\mu = e\indices{_a^\mu} \tilde{\gamma}^a,
\end{equation}
where curved and flat space $\gamma$-matrices are defined by
\begin{equation}
\{ \gamma^\mu, \gamma^\nu \} = 2g^{\mu\nu}, 
\qquad
\{ \tilde{\gamma}^a, \tilde{\gamma}^b \} = 2  \eta^{ab},
\end{equation}
where $\{x,y\} \equiv xy + yx$ is the anticommutator.  From the vierbein, one can construct the spin connection,
\begin{align}
w_{\mu ab} &= 
\frac{1}{2} e\indices{_a^\alpha} (\partial_\mu e_{b\alpha} - \partial_\alpha e_{b\mu} )
+ \frac{1}{2} e\indices{_b^\beta}
(\partial_\beta e_{a\mu} - \partial_\mu e_{a\beta} )
\notag
\\
&\qquad - \frac{1}{2} 
e\indices{^c_\mu}
e\indices{_a^\alpha}
e\indices{_b^\beta}  (\partial_\alpha e_{c\beta} - \partial_\beta e_{c\alpha} ),
\end{align}
and then the spinor connection,
\begin{equation}
\Gamma_\mu = - \frac{1}{4} \tilde{\gamma}^a \tilde{\gamma}^b \omega_{\mu a b}.
\end{equation}
With the spinor connection, we can define the covariant derivative of spinors,
\begin{equation}
\nabla_\mu \psi = \partial_\mu \psi - \Gamma_\mu \psi,
\qquad
\nabla_\mu \bar{\psi} = \partial_\mu \bar{\psi} + \bar{\psi} \Gamma_\mu.
\end{equation} 

In our framework, we include a single Dirac field in the matter sector.  This is notably different from previous studies of the spherically symmetric Einstein-Dirac system, in which at least two Dirac fields are included.  The inclusion of two fields is explained as being necessary for preserving spherical symmetry.  More generally, $2j+1$ fields are included when the total angular momentum of the system is given by $j$.  We stress that, in these previous studies, these fields were never quantized.  In our framework, we quantize a single field and construct spherically symmetric configurations through excitations of the vacuum.

From the Lagrangian in (\ref{Lagrangian}), we can derive the classical stress-energy-momentum tensor,
\begin{equation} \label{energy-momentum}
\begin{split}
T_{\mu\nu} &= 
- \frac{1}{4}
\Bigl[
\bar{\psi} \gamma_\mu \nabla_\nu \psi
+ \bar{\psi} \gamma_\nu \nabla_\mu \psi
\\
&\qquad
- (\nabla_\mu \bar{\psi})\gamma_\nu  \psi
- (\nabla_\nu \bar{\psi})\gamma_\mu \psi
\Bigr],
\end{split}
\end{equation}
and the classical equations of motion,
\begin{equation} \label{eom}
\begin{split}
0 &= \gamma^\mu \nabla_\mu \psi - m_\psi \psi
\\
&= \gamma^\mu (\partial_\mu - \Gamma_\mu) \psi - m_\psi \psi,
\end{split}
\end{equation}
which is the Dirac equation.

Given a Dirac spinor $\psi$, the charge conjugated spinor is given by
\begin{equation}
\psi^c \equiv \mathcal{C} \bar{\psi}^T,
\end{equation}
where $\mathcal{C}$ is the charge conjugation operator defined by $\mathcal{C}^T = -\mathcal{C}$ and $\mathcal{C} \tilde{\gamma}^{aT} \mathcal{C}^{-1} = -\tilde{\gamma}^a$.  It is straightforward to show that if $\psi$ satisfies the Dirac equation, $\psi^c$ does as well.  If we were to couple the Dirac field to the electromagnetic field, one can show that if $\psi$ satisfies the charged Dirac equation, then $\psi^c$ also satisfies the charged Dirac equation, but with an opposite sign for its charge.  In flat space, this indicates that if $\psi$ describes a particle, $\psi^c$ describes the antiparticle.  This interpretation continues to hold in curved space for the static metric in (\ref{metric}).

When performing calculations, we will use the Dirac representation for the flat space $\gamma$-matrices,
\begin{equation} \label{Dirac rep}
\tilde{\gamma}^0 = i
\begin{pmatrix}
1 & 0 \\ 0 & -1
\end{pmatrix},
\qquad
\tilde{\gamma}^j = i
\begin{pmatrix}
0 & \sigma^j \\ -\sigma^j & 0
\end{pmatrix},
\end{equation}
where $j=1,2,3$ and where the $\sigma^j$ are the Pauli matrices,
\begin{equation}
\sigma^1 = 
\begin{pmatrix}
0 & 1 \\ 1 & 0
\end{pmatrix},
\quad
\sigma^2 = 
\begin{pmatrix}
0 & -i \\ i & 0
\end{pmatrix}, 
\quad
\sigma^3 = 
\begin{pmatrix}
1 & 0 \\ 0 & -1
\end{pmatrix}.
\end{equation} 
Our convention for the adjoint spinor is
\begin{equation}
\bar{\psi} = \psi^\dag (-i\tilde{\gamma}^0).
\end{equation}
The charge conjugated spinor is then given by
\begin{equation} \label{psi c}
\psi^c = \tilde{\gamma}^2 \psi^*,
\end{equation}
where $\mathcal{C} = i\tilde{\gamma}^0 \tilde{\gamma}^2$.

We will also make a choice for the vierbein.  There are two standard choices available.  The two choices lead to identical radial equations of motion, but to different, although of course equivalent, descriptions for the angular part of the Dirac field.  One choice is
\begin{equation} \label{vierbein}
\gamma^t = \frac{\tilde{\gamma}^t}{\alpha}, \quad
\gamma^r = \frac{\tilde{\gamma}^r}{a}, \quad
\gamma^\theta = \frac{\tilde{\gamma}^\theta}{r}, \quad
\gamma^\phi = \frac{\tilde{\gamma}^\phi}{r\sin\theta},
\end{equation}
which gives the vierbein through (\ref{gamma e gamma}), where
\begin{equation} \label{gamma t r t p}
\begin{split}
\tilde{\gamma}^t &= \tilde{\gamma}^0
\\
\tilde{\gamma}^r &= \tilde{\gamma}^1 \sin\theta \cos\phi + \tilde{\gamma}^2 \sin\theta \sin\phi + \tilde{\gamma}^3 \cos\theta
\\
\tilde{\gamma}^\theta &= \tilde{\gamma}^1 \cos\theta \cos\phi + \tilde{\gamma}^2 \cos\theta \sin\phi - \tilde{\gamma}^3 \sin\theta
\\
\tilde{\gamma}^\phi &= -\tilde{\gamma}^1\sin\phi + \tilde{\gamma}^2\cos\phi .
\end{split}
\end{equation}
For this vierbein, the spinor connection works out to
\begin{align}
\Gamma_t &= \frac{\partial_r \alpha}{2a} \tilde{\gamma}^t
\tilde{\gamma}^r
\notag\\
\Gamma_r &= 0
\notag\\
\Gamma_\theta &= \frac{1}{2} \left(1- \frac{1}{a} \right)
\tilde{\gamma}^\theta \tilde{\gamma}^r
\notag\\
\Gamma_\phi &= \frac{1}{2} \left(1- \frac{1}{a} \right)
\sin\theta \,
\tilde{\gamma}^\phi\tilde{\gamma}^r ,
\label{spinor connection}
\end{align}
from which
\begin{equation} \label{gamma Gamma}
\gamma^\mu \Gamma_\mu = \frac{1}{a} \left( \frac{a}{r} - \frac{\partial_r \alpha}{2\alpha} - \frac{1}{r} \right) \tilde{\gamma}^r,
\end{equation}
which is needed for the Dirac equation in (\ref{eom}).  A benefit of using this vierbein is that it leads to solutions to the equations of motion that are similar to the standard solutions for the flat space Dirac equation in spherical coordinates \cite{Sakurai, Bransden}.  Further, angular momentum is described through the familiar functions of spherical harmonics and two-component spinors.

The other choice for the vierbein is
\begin{equation} \label{alt vierbein}
\gamma^t = \frac{\tilde{\gamma}^0}{\alpha}, \quad
\gamma^r = \frac{\tilde{\gamma}^3}{a}, \quad
\gamma^\theta = \frac{\tilde{\gamma}^2}{r}, \quad
\gamma^\phi = \frac{\tilde{\gamma}^1}{r\sin\theta}.
\end{equation}
Note that this vierbein is not diagonal.  By associating the off-diagonal Pauli matrices $\sigma^1$ and $\sigma^2$ with the angular $\gamma$-matrices $\gamma^\theta$ and $\gamma^\phi$, it becomes straightforward to introduce the raising and lowering operators for spin weighted spherical harmonics.

We find that solving the equations of motion is simpler using the vierbein in (\ref{alt vierbein}).  However, intuition for quantum angular momentum is typically based on the properties of spherical harmonics and two-component spinors, not on spin weighted spherical harmonics.  For this reason we use the vierbein in (\ref{vierbein}).  For completeness, we also solve the Dirac equation using the alternative choice in (\ref{alt vierbein}) in Appendix \ref{sec:alt eom}.


\section{Quantum Dirac spinors in curved spacetime}
\label{sec:quantize}

In this section, we canonically quantize the Dirac field in a static spherically symmetric curved spacetime background.  Quantum theories live in Hilbert space, i.e.~a complex vector space with an inner product.  We therefore begin by identifying an inner product.  First, we note that the current
\begin{equation}
j^\mu = i\bar{\psi}_1 \gamma^\mu \psi_2
\end{equation}
can be shown to be divergence-free, $\nabla_\mu j^\mu = 0$, with the help of the equations of motion in (\ref{eom}).  We can then define an inner product on the space of solutions to the Dirac equation,
\begin{equation} \label{inner product}
(\psi_1, \psi_2) \equiv \int_{\Sigma_t} d^3x \sqrt{\det(\gamma_{ij})}\, n_\mu j^\mu.
\end{equation}
We shall assume that $\psi_1$ and $\psi_2$ decay sufficiently quickly at spatial infinity.  As a consequence, this inner product is independent of the hypersurface $\Sigma_t$ over which the integral is evaluated.  Moving to the static spherically symmetric metric in (\ref{metric}) and to the vierbein in (\ref{vierbein}), the inner product can be written as
\begin{equation} \label{inner product 2}
(\psi_1, \psi_2) = \int_\Sigma d^3x \sqrt{\det(\gamma_{ij})}\, \psi_1^\dag \psi_2,
\end{equation}
where $\det(\gamma_{ij}) = a^2 r^4 \sin^2\theta$ and $d^3x = dr d\theta d\phi$.  The inner product is linear in both arguments and obeys the identities
\begin{equation}
\begin{split}
(\psi_1, \psi_2)^* &= (\psi_2, \psi_1)
\\
(\psi_1^c, \psi_2^c) &= (\psi_2, \psi_1)
\\
(\psi_1, \psi_2^c) &= (\psi_2, \psi_1^c).
\end{split}
\end{equation}

The momentum conjugate to $\psi$ is given by
\begin{equation} \label{pi def}
\pi = \frac{\partial \mathcal{L}}{\partial(\partial_t \psi)} = -\sqrt{\det(\gamma_{ij})} \, \bar{\psi} n_\mu \gamma^\mu
= \sqrt{\det(\gamma_{ij})} \, i \psi^\dag,
\end{equation}
where in the final equality we used the metric in (\ref{metric}) and the vierbein in (\ref{vierbein}).  We now promote $\psi$ and $\pi$ to operators $\hat{\psi}$ and $\hat{\pi}$ and impose the equal time anticommutation relations,
\begin{equation} \label{canonical anticommuation}
\begin{split}
\{ \hat{\psi}_a(t,\vec{x}), \hat{\pi}_b(t,\vec{y}) \} &= i\delta_{ab} \delta^3(\vec{x} - \vec{y}),
\\
\{ \hat{\psi}_a(t,\vec{x}), \hat{\psi}_b(t,\vec{y}) \} &= 0
\\
\{ \hat{\pi}_a(t,\vec{x}), \hat{\pi}_b(t,\vec{y}) \} &= 0,
\end{split}
\end{equation}
where $a,b$ label the Dirac spinor components.

Let $f_I$ be a set of classical solutions to the equations of motion,
\begin{equation}
(\gamma^\mu \nabla_\mu - m_\psi) f_I = 0,
\end{equation}
where the subscript $I$ represents some set of quantum numbers.  Since the static metric in (\ref{metric}) is time independent, it possesses the hyperspace-orthogonal timelike Killing vector $\xi^\mu = (1,0,0,0)$.  The existence of this vector allows for a coordinate invariant definition of positive frequency solutions \cite{Birrell:1982ix, Carroll:2004st},
\begin{equation} \label{pos sol}
\xi^\mu \partial_\mu f_I = \partial_t f_I =  -i\omega_I f_I,
\end{equation} 
and negative frequency solutions,
\begin{equation} \label{neg sol}
\xi^\mu \partial_\mu f_I = \partial_t f_I =  +i\omega_I f_I,
\end{equation}
for real positive $\omega_I$.  Labeling the positive frequency solutions with a $+$ and the negative frequency solutions with a $-$, we assume that the $f_I^\pm$ are orthonormal in the sense that
\begin{equation} \label{orthonormal}
\begin{split}
(f_I^+, f_J^+) &= (f_I^-, f_J^-) = \delta_{IJ}
\\
(f_I^+, f_J^-) &= (f_I^-, f_J^+) = 0.
\end{split}
\end{equation}
The $f_I^\pm$ can then be used as mode functions in the expansion of $\hat{\psi}$,
\begin{equation} \label{field decomposition}
\hat{\psi}(t,\vec{x}) = \sum_I \left[ \hat{b}_I f_I^+(t, \vec{x}) + \hat{d}^\dag_I f_I^-(t,\vec{x}) \right],
\end{equation}
which defines creation and annihilation operators.  The fact that we can classify mode functions as having positive or negative frequency allows for a natural definition of particles and antiparticles and we can interpret the corresponding creation and  annihilation operators in the field decomposition in (\ref{field decomposition}) as being able to create and annihilate particles and antiparticles.  We may solve for the creation and annihilation operators using the inner product.  Specifically,
\begin{equation} \label{particle operators}
\hat{b}_I^\dag = (\hat{\psi}, f_I^+),\qquad
\hat{b}_I = (f_I^+, \hat{\psi})
\end{equation}
are creation and annihilation operators for particles and 
\begin{equation} \label{antiparticle operators}
\hat{d}_I^\dag = (f_I^-, \hat{\psi}), \qquad
\hat{d}_I = (\hat{\psi}, f_I^-)
\end{equation}
are creation and annihilation operators for antiparticles.  

Using Eqs.~(\ref{particle operators}) and (\ref{antiparticle operators}), the definition of $\pi$ in (\ref{pi def}), and the canonical anticommutation relations in (\ref{canonical anticommuation}), we can derive
\begin{equation} \label{b d anticommutation}
\begin{split}
\{ \hat{b}_I, \hat{b}^\dag_J \} &= (f_I^+, f_J^+)
\\
\{ \hat{d}^\dag_I, \hat{d}_J \} &= (f_I^-, f_J^-)
\\
\{\hat{b}_I, \hat{d}_J \} &= (f_I^+, f_J^-)
\\
\{\hat{d}_I^\dag, \hat{b}_J^\dag \} &= (f_I^-, f_J^+),
\end{split}
\end{equation}
with all other anticommutators vanishing.  Since we are assuming our mode functions satisfy the orthonormality relations in (\ref{orthonormal}), then the anticommutation relations in (\ref{b d anticommutation}) reduce down to standard anticommutation relations for creation and annihilation operators for a Dirac field.


\section{Solutions to the classical equations of motion}
\label{sec:eom sol}

Our goal in this section is to solve the classical equations of motion, i.e.~the Dirac equation,
\begin{equation} \label{eom 2}
\left[\gamma^\mu (\partial_\mu - \Gamma_\mu) - m_\psi \right] f_I = 0,
\end{equation}
using the vierbein in (\ref{vierbein}), for the mode functions $f_I$ and to show that the mode functions satisfy the orthonormality conditions in (\ref{orthonormal}) and the positive and negative frequency conditions in (\ref{pos sol}) and (\ref{neg sol}).  Over the course of doing this, we will determine the precise quantum numbers that are included in the subscript $I$.  For completeness, we also solve the equations of motion using the alternative choice for the vierbein in (\ref{alt vierbein}) in Appendix \ref{sec:alt eom}.

To write the equations of motion in a convenient form, we recall the standard quantum operators for orbital angular momentum \cite{Sakurai, Bransden},
\begin{equation} \label{L operators}
\begin{split}
\widehat{L}_1 &= -i \left( -\sin\phi \,\partial_\theta - \cot\theta \cos\phi \, \partial_\phi \right)
\\
\widehat{L}_2 &= -i \left(+ \cos\phi \,\partial_\theta - \cot\theta \sin\phi \, \partial_\phi \right)
\\
\widehat{L}_3 &= -i \partial_\phi.
\end{split}
\end{equation}
We introduce the spin operator for a four-component Dirac spinor,
\begin{equation}
\widehat{\mathbf{S}} = \frac{1}{2} \widehat{\mathbf{\Sigma}},
\end{equation}
where
\begin{equation} \label{Sigma def}
\widehat{\Sigma}_1 = 
\begin{pmatrix}
\sigma^1 & 0 \\ 0 & \sigma^1
\end{pmatrix}
\quad
\widehat{\Sigma}_2 = 
\begin{pmatrix}
\sigma^2 & 0 \\ 0 & \sigma^2
\end{pmatrix}
\quad
\widehat{\Sigma}_3 = 
\begin{pmatrix}
\sigma^3 & 0 \\ 0 & \sigma^3
\end{pmatrix}.
\end{equation}
Using the Dirac representation for $\gamma$-matrices in (\ref{Dirac rep}), $\widehat{\Sigma}_j$ can be written as
\begin{equation} \label{Sigma j}
\widehat{\Sigma}_j = -i \tilde{\gamma}^j \tilde{\gamma}^1 \tilde{\gamma}^2 \tilde{\gamma}^3.
\end{equation}
With $\widehat{\mathbf{L}}$ and $\widehat{\mathbf{\Sigma}}$, we define the operator
\begin{equation} \label{K def}
\widehat{K} \equiv -i \tilde{\gamma}^0 \left(1 + \widehat{\mathbf{L}} \cdot \widehat{\mathbf{\Sigma}}\right),
\end{equation}
where $\widehat{\mathbf{L}} \cdot \widehat{\mathbf{\Sigma}} = \widehat{L}_1 \widehat{\Sigma}_1 + \widehat{L}_2 \widehat{\Sigma}_2 + \widehat{L}_3 \widehat{\Sigma}_3$.  The quantum operator for total angular momentum is given as usual by
\begin{equation}
\widehat{\mathbf{J}} = \widehat{\mathbf{L}} + \widehat{\mathbf{S}},
\end{equation}
and $\widehat{K}$ can also be written as
\begin{equation} \label{K J L}
\widehat{K} = -i\tilde{\gamma}^0 \left(\widehat{J}^{\,2} - \widehat{L}^{2} + \frac{1}{4}\right).
\end{equation}

Returning to the equations of motion in (\ref{eom 2}), we allow $f$ to have arbitrary dependencies,
\begin{equation}
f = f(t,r,\theta,\phi),
\end{equation}
and, for the moment, suppress the subscripted $I$.  Using the vierbein in (\ref{vierbein}) and Eqs.~(\ref{gamma Gamma}) and (\ref{K def}), the equations of motion can be written
\begin{equation} \label{H f t eom}
\widehat{H} f = i\partial_t f,
\end{equation}
where
\begin{equation}  \label{Hhat def}
\widehat{H} \equiv \frac{i\alpha}{a} \tilde{\gamma}^0 \tilde{\gamma}^r
\left(\partial_r  + \frac{\partial_r \alpha}{2\alpha}  + \frac{1}{r} \right)
- \frac{\alpha}{r} \tilde{\gamma}^r \widehat{K}
- i\alpha m_\psi \tilde{\gamma}^0.
\end{equation}
One can show that the operators $i\partial_t$, $\widehat{H}$, $\widehat{J}^{\,2}$, $\widehat{J}_3$, and $\widehat{K}$ commute with one another.  This is somewhat tedious to do, but the calculations are aided by first showing that $\tilde{\gamma}^r$ commutes with $\widehat{K}$ and $\widehat{\mathbf{J}}$.  Since all five operators commute with one another, we can assume $f$ is a simultaneous eigenfunction of them.  

For $i\partial_t$ we have the eigenvalue equation
\begin{equation}
i\partial_t f = \omega f.
\end{equation}
It is not difficult to show that $i\partial_t$ is Hermitian with respect to the inner product in Eq.\ (\ref{inner product 2}),
\begin{equation}
(f,i\partial_t f) = (i\partial_t f, f), 
\end{equation}
as long as $f$ decays sufficiently quickly at spatial infinity, and hence $\omega$ is real.  We now assume that the time dependence of $f$ is separable, so that $f$ can be written as
\begin{equation} \label{f omega u}
f(t,r,\theta,\phi) = e^{-i\omega t} u(r,\theta,\phi).
\end{equation}
The equations of motion can now be written
\begin{equation} \label{H u eom}
\widehat{H}u = \omega u.
\end{equation}
We have also
\begin{equation} \label{K u eq}
\widehat{K} u = \kappa u,
\end{equation} 
where $\kappa$ are the eigenvalues of $\widehat{K}$.

The next step in our solution is to write $u$ as
\begin{equation}
u(r,\theta,\phi) = 
\begin{pmatrix}
\chi(r,\theta,\phi) \\ \eta(r,\theta,\phi)
\end{pmatrix},
\end{equation}
where $\chi$ and $\eta$ are two-component objects.  Plugging this into Eq.~(\ref{K u eq}) and using the definition of $\widehat{K}$ in (\ref{K def}), we find the two equations
\begin{equation} \label{xi pm eval eqs}
\begin{split} \
(1 + \widehat{\mathbf{L}} \cdot \vec{\sigma} ) \xi &= +\kappa \chi
\\
(1 + \widehat{\mathbf{L}} \cdot \vec{\sigma} ) \eta &= -\kappa \eta,
\end{split}
\end{equation}
where $\widehat{\mathbf{L}} \cdot \vec{\sigma} = \widehat{L}_1\sigma^1 + \widehat{L}_2\sigma^2 + \widehat{L}_3 \sigma^3$.  The eigenfunctions and eigenvalues of $1+\widehat{\mathbf{L}} \cdot \vec{\sigma}$ are well known \cite{Sakurai, Bransden},
\begin{equation} \label{chi evalue eq}
(1 + \widehat{\mathbf{L}} \cdot \vec{\sigma} ) \mathcal{Y}_{j\pm 1/2}^{m_j} = \mp (j+1/2) \mathcal{Y}_{j\pm 1/2}^{m_j},
\end{equation}
where the $\mathcal{Y}_{\ell}^{m_j}$ are the spin-angle functions
\begin{equation} \label{spin angle}
\begin{split}
\mathcal{Y}_{j-1/2}^{m_j} &= 
\sqrt{\frac{j+m_j}{2j}} Y_{j-1/2}^{m_j-1/2}
\begin{pmatrix}
1 \\ 0
\end{pmatrix}
\\
&\qquad
+
\sqrt{\frac{j-m_j}{2j}} Y_{j-1/2}^{m_j+1/2}
\begin{pmatrix}
0 \\ 1
\end{pmatrix}
\\
\mathcal{Y}_{j+1/2}^{m_j} &= 
- \sqrt{\frac{j+1-m_j}{2j+2}} Y_{j+1/2}^{m_j-1/2}
\begin{pmatrix}
1 \\ 0
\end{pmatrix}
\\
&\qquad
+
\sqrt{\frac{j+1+m_j}{2j+2}} Y_{j+1/2}^{m_j+1/2}
\begin{pmatrix}
0 \\ 1
\end{pmatrix},
\end{split}
\end{equation}
where the $Y_\ell^{m_\ell}$ are spherical harmonics.  We can see that $\mathcal{Y}_{j-1/2}^{m_j}$ is the linear combination of products of spherical harmonics with orbital angular momentum $\ell = j-1/2$ and two-component spinors with spin angular momentum $s = 1/2$ with the appropriate Clebsch-Gordon coefficients such that the total angular momentum is $j$ and the total angular momentum 3-component is $m_j$.  An identical statement can be made about $\mathcal{Y}_{j+1/2}^{m_j}$, except $\ell = j+1/2$.  Since total angular momentum is the sum of $\ell$ and $s=1/2$, $j=1/2, 3/2, 5/2,\ldots$ and $m_j = -j, -j+1, \ldots, j-1,j$.  Comparing Eqs.~(\ref{xi pm eval eqs}) and (\ref{chi evalue eq}), we have found the eigenvalues of $\widehat{K}$,
\begin{equation} \label{lambda eval}
\kappa = \pm(j+1/2).
\end{equation}

We now assume that the radial ($r$) and angular ($\theta,\phi$) dependence separates.  Again comparing Eqs.~(\ref{xi pm eval eqs}) and (\ref{chi evalue eq}), we can see that the eigenfunctions to Eq.~(\ref{K u eq}) are
\begin{equation} \label{u pm}
u_\pm(r,\theta,\phi) =  
\begin{pmatrix}
R_\pm^{(1)}(r) \mathcal{Y}_{j\mp 1/2}^{m_j}(\theta,\phi)
\\[6pt]
R_\pm^{(2)}(r) \mathcal{Y}_{j\pm 1/2}^{m_j}(\theta,\phi)
\end{pmatrix},
\end{equation}
where $R_\pm^{(1)}$ and $R_\pm^{(2)}$ are four arbitrary one-component functions and where the upper/lower signs are consistent with those for the eigenvalues in (\ref{lambda eval}).

It is not difficult to show that $u_\pm$ are eigenfunctions of $\widehat{J}^{\,2}$ and $\widehat{J}_3$,
\begin{equation}
\begin{split}
\widehat{J}^{\,2} u_\pm = j(j+1)u_\pm,
\qquad
\widehat{J}_3 u_\pm = m_j u_\pm.
\end{split}
\end{equation}
That they are eigenfunctions of $\widehat{J}_3$ is easy to see using $\widehat{J}_3 = \widehat{L}_3 + \widehat{S}_3$ and the standard formulas for $\widehat{L}_3$ and $\widehat{S}_3$ acting on their respective eigenfunctions.  That they are eigenfunctions of $\widehat{J}^{\,2}$ can be shown by writing  $\widehat{J}^{\,2}$ in terms of $\widehat{L}^2$ and the operator in Eq.~(\ref{chi evalue eq}).  We have now established that $u_\pm$ in (\ref{u pm}) are simultaneous eigenfunctions of $\widehat{H}$, $\widehat{K}$, $\widehat{J}^{\,2}$, and $\widehat{J}_3$.

Before turning to the radial equations of motion, we outline the derivation of a useful result.  We previously mentioned that $\tilde{\gamma}^r$ commutes with $\widehat{K}$.  This means that $\tilde{\gamma}^r u_\pm$ are eigenfunctions of $\widehat{K}$ since
\begin{equation}
\pm(j+1/2) \tilde{\gamma}^r u_\pm 
= \tilde{\gamma}^r\widehat{K} u_\pm
= \widehat{K} \tilde{\gamma}^r u_\pm .
\end{equation}
Using this, and that $\tilde{\gamma}^r$ can be written as
\begin{equation}
\tilde{\gamma}^r = 
i
\begin{pmatrix}
0 & \sigma^r \\
-\sigma^r & 0
\end{pmatrix},
\end{equation}
where
\begin{equation}
\sigma^r \equiv \sigma^1 \sin\theta \cos\phi + \sigma^2 \sin\theta \sin\phi + \sigma^3 \cos\theta,
\end{equation}
one can show that 
\begin{equation}
\sigma^r \mathcal{Y}_{j\pm1/2}^{m_j} = \mathcal{Y}_{j\mp 1/2}^{m_j}
\end{equation}
and thus that
\begin{equation} \label{gamma r spinor}
\tilde{\gamma}^r u_\pm = \tilde{\gamma}^r
\begin{pmatrix}
R_\pm^{(1)} \mathcal{Y}_{j\mp 1/2}^{m_j}
\\[6pt]
R_\pm^{(2)} \mathcal{Y}_{j\pm 1/2}^{m_j}
\end{pmatrix}
= i
\begin{pmatrix}
R_\pm^{(2)} \mathcal{Y}_{j\mp1/2}^m \\ - R_\pm^{(1)} \mathcal{Y}_{j\pm 1/2}^m
\end{pmatrix}.
\end{equation}

We now plug the results derived so far into the equations of motion in (\ref{H u eom}).  We find that the angular dependence cancels out, leaving us with the radial equations of motion
\begin{equation} \label{radial eom pm}
\begin{split}
\omega R_\pm^{(2)}
&= 
-\frac{i\alpha}{a} 
\left(\partial_r  + \frac{\partial_r \alpha}{2\alpha} + \frac{1}{r} \right) R_\pm^{(1)}
- \alpha m_\psi R_\pm^{(2)}
\\
&\qquad 
\pm \frac{i\alpha}{r} \left(j+\frac{1}{2}\right) R_\pm^{(1)}
\\
\omega R_\pm^{(1)}
&= 
-\frac{i\alpha}{a} 
\left(\partial_r  + \frac{\partial_r \alpha}{2\alpha} + \frac{1}{r} \right) R_\pm^{(2)}
+ \alpha m_\psi R_\pm^{(1)}
\\
&\qquad
\mp \frac{i\alpha}{r}\left(j+\frac{1}{2}\right) R_\pm^{(2)}.
\end{split}
\end{equation}

We have found that solutions are distinguished by the quantum numbers $j$, $m_j$, and $\pm$, where $j=1/2, 3/2, 5/2, \ldots$, $m_j = -j, -j+1, \ldots, j-1,j$, and $\pm$ is the sign of the eigenvalue $\kappa$.  In Sec.~\ref{ED SG}, we will find that the outer boundary conditions we impose when solving the radial equations of motion lead to an additional quantum number, $n = 0,1,2,\ldots$, which equals the numbers of nodes in the solution to the radial equations of motion.  For our solutions to the classical equations of motion, $f_I$, we have $I = \{n,j,m_j,\pm\}$.  The radial equations in (\ref{radial eom pm}) are independent of $m_j$ and we can therefore take their solutions to also be independent of $m_j$:~$R^{(1)}_{nj\pm}$, $R^{(2)}_{nj\pm}$, and $\omega_{nj\pm}$.

At this point, we recall that if $f_I$ is a solution to the equations of motion, then the charge conjugated $f_I^c$ is also a solution to the equations of motion.  From (\ref{psi c}) and (\ref{f omega u}),
\begin{equation}
f_I^c = e^{+i\omega_I t} u_I^c,
\end{equation}
where $u_I^c = \tilde{\gamma}^2 u_I^*$.  Plugging this into (\ref{H f t eom}), we have
\begin{equation} \label{H uc eom}
\widehat{H} u_I^c = - \omega u_I^c.
\end{equation}
We can also derive this by multiplying Eq.~(\ref{H u eom}) by $\tilde{\gamma}^2$ from the left and noting that $(\tilde{\gamma}^2 \tilde{\gamma}^0)^* = \tilde{\gamma}^0 \tilde{\gamma}^2$, $(\tilde{\gamma}^2 \tilde{\gamma}^r)^* = \tilde{\gamma}^r \tilde{\gamma}^2$, and $(\tilde{\gamma}^2 \widehat{K})^* = - \widehat{K}\tilde{\gamma}^2$.  Comparing (\ref{H uc eom}) to (\ref{H u eom}), we find that if $u_I$ solves the equations of motion with eigenvalue $\omega_I$, then $u_I^c$ solves the equations of motion with eigenvalue $-\omega_I$.

We expect the general form for our solution in Eq.~(\ref{u pm}) to accommodate both $u_I$ and $u_I^c$.  The charge conjugation of Eq.~(\ref{u pm}) works out to
\begin{equation}
u_{njm_j\pm}^c = \tilde{\gamma}^2 u_{njm_j\pm}^*
= (-1)^{m\mp1/2}
\begin{pmatrix}
R_{nj\pm}^{(2)*} \mathcal{Y}_{j\pm 1/2}^{-m_j}
\\[6pt]
R_{nj\pm}^{(1)*} \mathcal{Y}_{j\mp 1/2}^{-m_j}
\end{pmatrix}.
\end{equation}
Dropping the irrelevant global phase and comparing $f_{njm_j\pm}$ with $f^c_{n,j,-m_j,\mp}$, we can identify
\begin{equation} \label{identification}
\omega_{nj\pm} = -\omega_{nj\mp}, \qquad
R^{(1)}_{nj\pm} = R^{(2)*}_{nj\mp}.
\end{equation}
It is straightforward to show that Eq.~(\ref{identification}) is a symmetry of the radial equations of motion in (\ref{radial eom pm}).

We are nearly ready to write down the final form for our solutions to the classical equations of motion.  The radial equations of motion simplify when written in terms of
\begin{equation}
\begin{split}
P_{nj+}(r) \equiv \hphantom{i} &r \sqrt{\alpha(r)} \, R_{nj+}^{(1)}(r)
\\
P_{nj-}(r) \equiv i &r \sqrt{\alpha(r)} \, R_{nj+}^{(2)*}(r).
\end{split}
\end{equation}
We include the factor of $i$ in the definition of $P_{nj-}$ because it leads to the cancellation of all $i$s in the radial equations of motion.  The final form for our solutions is then
\begin{equation} \label{mode functions}
f_{njm_j\pm} (t,r,\theta,\phi)
=
\frac{e^{\mp i\omega_{nj} t}}{r\sqrt{\alpha(r)}}
\begin{pmatrix}
\hphantom{i}P_{nj\pm} (r)
\mathcal{Y}_{j\mp 1/2}^{m_j} (\theta,\phi)\\[4pt]
iP^*_{nj\mp} (r)
\mathcal{Y}_{j\pm1/2}^{m_j} (\theta,\phi)
\end{pmatrix}.
\end{equation}
We no longer include the $\pm$ in the subscript of $\omega_{nj}$ and take $\omega_{nj}$ to be positive.  It is easily seen that the mode functions in (\ref{mode functions}) satisfy the positive and negative frequency conditions in (\ref{pos sol}) and (\ref{neg sol}), with the positive and negative frequency solutions being labeled by $\pm$.  We have not included a normalization constant because the method we use to normalize the mode functions, which we present in Sec.~\ref{sec:numerical}, does not require one.  Finally, we note that $f^c_{njm_j+} = i (-1)^{m_j+1/2} f_{n,j,-m_j,-}$, and hence that $f^c_{njm_j+}$ and $f_{n,j,-m_j,-}$ differ by an irrelevant global phase.

In terms of $P_{nj\pm}$, the radial equations of motion in (\ref{radial eom pm}) become
\begin{equation} \label{P radial eom}
\begin{split}
\partial_r P_{nj\pm} 
&= \mp\frac{a}{\alpha}
\left[ (\omega_{nj} \pm \alpha m_\psi) P^*_{nj\mp} - \frac{\alpha}{r} \left(j + \frac{1}{2}\right) P_{nj\pm} \right].
\end{split}
\end{equation}
As promised, all factors of $i$ have disappeared.  This suggests that the $P_{nj\pm}$ are real.  We will show in Sec.~\ref{sec:semiclassical gravity} that the $P_{nj\pm}$ being real leads to a diagonal stress-energy-momentum tensor and hence to a spherically symmetric and static spacetime.

Having found the solutions to the equations of motion, it remains to show that they satisfy the orthonormality conditions in (\ref{orthonormal}).  The spin-angle functions in (\ref{spin angle}) satisfy 
\begin{equation} \label{orthonormal spin angle}
\int d\theta d\phi \sin\theta
\left(\mathcal{Y}_{j\pm1/2}^{m_j}\right)^\dag
\mathcal{Y}_{j'\pm'1/2}^{m_j'}
= \delta_{j, j'} \delta_{m_j, m_j'} \delta_{\pm, \pm'},
\end{equation}
which can be shown using the standard orthonormality formulas for spherical harmonics and two-component spinors.  Plugging Eq.~(\ref{mode functions}) into the inner product in (\ref{inner product 2}) and using (\ref{orthonormal spin angle}), we have
\begin{equation} \label{fI fIp}
\begin{split}
&(f_{njm_j\pm}, f_{n'j'm_j'\pm'})
\\
&\qquad
= \delta_{j,j'} \delta_{m_j,m_j'} \delta_{\pm, \pm'}
e^{\pm i(\omega_{nj} - \omega_{n'j})t}
\\
&\qquad\qquad
\times\int_0^\infty dr \frac{a}{\alpha}
\left( 
P_{nj\pm}^* 
P_{n'j\pm}
+
P_{nj\mp} 
P^*_{n'j\mp}
\right).
\end{split}
\end{equation}
We find that orthogonality of the mode functions comes down to orthogonality with respect to the quantum number $n$.  We proceed as follows.    The mode functions may be written as $f_I = e^{-i\omega_I t} u_I$.  The $u_I$ are eigenfunctions of $\widehat{H}$ with eigenvalue $\omega_I$, $\widehat{H} u_I = \omega_I u_I$, where $\widehat{H}$ is given in (\ref{Hhat def}).  The inner product in (\ref{inner product 2}) becomes
\begin{equation} \label{f u product relation 0}
(f_I, f_{J}) 
= e^{\pm i(\omega_I - \omega_{J})t}
( u_{I}, u_{J} ).
\end{equation}
In Appendix \ref{sec:hermiticity}, we show that $\widehat{H}$ is Hermitian with respect to $( u_{I}, u_{J} )$,
\begin{equation}
( u_{I}, \widehat{H} u_{J} )
= ( \widehat{H} u_{I}, u_{J} ).
\end{equation}
As a consequence, mode functions are orthogonal with respect to the quantum number $n$ as long as they have distinct eigenvalues $\omega_{nj}$.  Finally, it follows from (\ref{fI fIp}) that our normalization requirement for a mode function is
\begin{equation} \label{norm req}
\mathcal{N}_{nj} \equiv (f_I, f_I) = \int_0^\infty dr \frac{a}{\alpha}
\left( 
|P_{nj+}|^2 
+
|P_{nj-} |^2
\right) 
= 1.
\end{equation}
Since the integrand is positive definite, it is possible to scale the $P_{nj\pm}$, and hence scale the $f_I$, to satisfy this normalization condition.  We explain the way in which we impose this normalization requirement in Sec.~\ref{sec:numerical}.  We have now shown that our mode functions in (\ref{mode functions}) satisfy the orthonormality conditions in (\ref{orthonormal}).  


\section{Spherically symmetric states}
\label{sec:states}

In Sec.~\ref{sec:quantize}, we canonically quantized the Dirac field.  We found creation and annihilation operators that obey the anticommutation relations in (\ref{b d anticommutation}).  Since our mode functions obey the orthonormality conditions in (\ref{orthonormal}), our anticommutation relations are the standard ones for a Dirac field.  The creation and annihilation operators may therefore be used to define a basis for the Hilbert space in the usual way.  Introducing the vacuum state $|0\rangle$, which is normalized and defined by
\begin{equation}
\hat{b}_I |0\rangle = \hat{d}_J |0\rangle = 0,
\end{equation}
basis states are constructed from repeated application of creation operators,
\begin{equation} \label{basis state}
\begin{split}
&|N_{I_1}^b, N_{I_2}^b, \ldots,N_{J_1}^d, N_{J_2}^d,\ldots\rangle
\\
&\qquad = 
\cdots (\hat{d}^\dag_{J_2})^{N_{J_2}^d} (\hat{d}^\dag_{J_1})^{N_{J_1}^d}
\cdots (\hat{b}^\dag_{I_2})^{N_{I_2}^b} (\hat{b}^\dag_{I_1})^{N_{I_1}^b} 
|0\rangle,
\end{split}
\end{equation}
where $N_I^b$ and $N_J^d$ are occupation numbers.  These states are normalized, orthogonal, and are eigenstates of the number operators,
\begin{equation}
\begin{split}
\widehat{N}_I^b |N_{I_1}^b, \ldots,N_{J_1}^d,\ldots\rangle
&= N_I^b |N_{I_1}^b, \ldots,N_{J_1}^d,\ldots\rangle
\\
\widehat{N}_J^d |N_{I_1}^b, \ldots,N_{J_1}^d,\ldots\rangle
&= N_J^d |N_{I_1}^b, \ldots,N_{J_1}^d,\ldots\rangle,
\end{split}
\end{equation}
where the number operators are given by
\begin{equation}
\widehat{N}_I^b = \hat{b}^\dag_I \hat{b}_I, \qquad
\widehat{N}_J^d = \hat{d}^\dag_J \hat{d}_J.
\end{equation}
Since the creation operators satisfy anticommutation relations, these states are totally antisymmetric in the nonzero occupation numbers and the occupation numbers can only equal $N^b_I, N_J^d=0$ or $1$.  A general state in the Hilbert space is an arbitrary linear combination of basis states,
\begin{equation} \label{general state}
|\psi\rangle = \sum_{N_I^b, N_J^d = 0}^1 C_{N_I^b, N_J^d}  
|N_{I_1}^b, N_{I_2}^b, \ldots,N_{J_1}^d, N_{J_2}^d,\ldots\rangle,
\end{equation}
such that the complex numbers $C_{N_I^b, N_J^d}$ satisfy $\sum_{N_I^b, N_J^d = 0}^1 |C_{N_I^b, N_J^d} |^2 = 1$.

Since we are restricting spacetime to be spherically symmetric, we must also restrict the basis states to be spherically symmetric.  Spherically symmetric states have zero total angular momentum \cite{Finster:1998ju, Olabarrieta:2007di, Alcubierre:2018ahf}.  For convenience, we focus on $b$-type excitations (which correspond to positive frequency mode functions).  Analogous results apply to $d$-type excitations.  An example of a spherically symmetric state is
\begin{equation} \label{zero j state}
|n,j,+\rangle \equiv \prod_{m_j=-j}^j \hat{b}^\dag_{njm_j} |0\rangle.
\end{equation}
That is, for fixed quantum numbers $n$ and $j$, all possible $m_j$ are excited.  We shall show that this state has zero total angular momentum, even though its individual excitations have nonzero angular momentum.  One can also excite multiple values of $n$ and $j$ as long as all possible $m_j$ are excited for each excited $j$.

Consider first the single-excitation state
\begin{equation} \label{single j mj}
|n,j,m_j,+\rangle \equiv \hat{b}^\dag_{njm_j} |0\rangle.
\end{equation}
Since this state has definite quantum number $m_j$, it must be an eigenstate of the $\widehat{J}_3$ operator,
\begin{equation}
\widehat{J}_3 |n,j,m_j,+\rangle = m_j |n,j,m_j,+\rangle.
\end{equation}
This implies that $\widehat{J}_3 = m_j \widehat{N}^b_{njm_j}$, from which
\begin{equation}
\widehat{J}_{3,\text{tot}} =\sum_I \widehat{J}_3 =  \sum_I m_j \widehat{N}^b_I,
\end{equation}
where $I = n,j,m_j$.  Applying this to the spherically symmetric state in (\ref{zero j state}),
\begin{equation}
\widehat{J}_{3,\text{tot}} |n,j,+\rangle
= \left(\sum_{m_j = -j}^j m_j \right) |n,j,+\rangle
= 0,
\end{equation}
and the 3-component of total angular momentum is zero.  It remains to show that the 1- and 2-components are also zero.

Next consider the ladder operators $\widehat{J}_\pm = \widehat{J}_1 \pm i \widehat{J}_2$, which change $m_j$ to $m_j\pm1$.  Applied to the single-excitation state in Eq.~(\ref{single j mj}), 
\begin{equation}
\begin{split}
&\widehat{J}_\pm |n,j,m_j,+\rangle
\\
&\qquad= \sqrt{j(j+1) - m_j(m_j\pm 1)} \, |n,j,m_j\pm 1,+\rangle
\\
&\qquad= \sqrt{j(j+1) - m_j(m_j\pm 1)} \, \hat{b}^\dag_{n, j,m_j\pm1} |0\rangle.
\end{split}
\end{equation}
This implies that 
\begin{equation}
\widehat{J}_\pm = \sqrt{j(j+1) - m_j(m_j\pm 1)} \, 
\hat{b}^\dag_{n,j,m_j\pm1} \hat{b}_{njm_j},
\end{equation}
from which
\begin{equation}
\widehat{J}_{1,\text{tot}} = \frac{1}{2}\sum_I \left( \widehat{J}_+ + \widehat{J}_- \right),
\quad
\widehat{J}_{2,\text{tot}} = \frac{1}{2i}\sum_I \left( \widehat{J}_+ - \widehat{J}_- \right),
\end{equation}
where $I = n,j,m_j$.  We now apply these to the spherically symmetric state in (\ref{zero j state}).  Since every value of $m_j$ is excited in the spherically symmetric state, an additional excitation must give zero by the Pauli exclusion principle and thus
\begin{equation}
\widehat{J}_{1,\text{tot}} |n,j,+\rangle 
= \widehat{J}_{2,\text{tot}} |n,j,+\rangle 
= 0.
\end{equation}
Since all components of $\widehat{\mathbf{J}}$ are zero for $|n,j,+\rangle$, this state has zero total angular momentum and is a spherically symmetric state.


\section{Semiclassical gravity}
\label{sec:semiclassical gravity}

In the semiclassical gravity approximation, the Einstein field equations are sourced by the expectation value of the stress-energy-momentum tensor \cite{Birrell:1982ix, Wald:1995yp, Parker:2009uva, Hu:2020luk},
\begin{equation} \label{EFE}
G_{\mu\nu} = 8\pi G \langle \widehat{T}_{\mu\nu} \rangle,
\end{equation}
where $G_{\mu\nu}$ is the Einstein tensor, $G$ is the gravitational constant, and $\langle \widehat{T}_{\mu\nu} \rangle = \langle \psi |\widehat{T}_{\mu\nu} |\psi\rangle$ for some state $|\psi\rangle$.  Gravity, as described by the left-hand side of Eq.~(\ref{EFE}), is treated classically, while a quantum description of matter is used in the right-hand side.

The classical stress-energy-momentum tensor for a Dirac spinor field was given in (\ref{energy-momentum}).  Promoting this to an operator gives
\begin{equation} \label{hat T}
\begin{split}
\widehat{T}_{\mu\nu} &= 
- \frac{1}{4}
\Bigl[
\hat{\bar{\psi}} \gamma_\mu \nabla_\nu \hat{\psi}
+ \hat{\bar{\psi}} \gamma_\nu \nabla_\mu \hat{\psi}
\\
&\qquad
- (\nabla_\mu \hat{\bar{\psi}})\gamma_\nu  \hat{\psi}
- (\nabla_\nu \bar{\hat{\psi}})\gamma_\mu \hat{\psi}
\Bigr].
\end{split}
\end{equation}
Inserting the field decomposition for $\hat{\psi}$ in (\ref{field decomposition}), we find
\begin{equation} \label{That b d}
\begin{split}
\widehat{T}_{\mu\nu} &= \sum_{I,J} \Bigl[
T_{\mu\nu}(f_I^+, f_J^+) \hat{b}^\dag_I \hat{b}_J 
+ T_{\mu\nu}(f_I^+, f_J^-) \hat{b}^\dag_I \hat{d}^\dag_J 
\\
&\qquad
+ T_{\mu\nu}(f_I^-, f_J^+)\hat{d}_I \hat{b}_J 
+ T_{\mu\nu}(f_I^-, f_J^-)\hat{d}_I \hat{d}^\dag_J 
\Bigr],
\end{split}
\end{equation}
where
\begin{equation} \label{T(fI, fJ)}
\begin{split}
T_{\mu\nu}(f_I, f_J) &\equiv -\frac{1}{4} \Bigl[
\bar{f}_I \gamma_\mu \nabla_\nu f_J
+ \bar{f}_I \gamma_\nu \nabla_\mu f_J
\\
&\qquad
- (\nabla_\mu \bar{f}_I) \gamma_\nu f_J
- (\nabla_\nu \bar{f}_I) \gamma_\mu f_J
\Bigr].
\end{split}
\end{equation}
Since the stress-energy-momentum tensor contains products of the field operator, products of the creation and annihilation operators are found in Eq.~(\ref{That b d}).  This leads to the expectation value of the stress-energy-momentum tensor being divergent, which must be handled through a regularization and renormalization procedure \cite{Birrell:1982ix, Wald:1995yp, Parker:2009uva}.  Following \cite{Alcubierre:2022rgp}, in this initial work on the Einstein-Dirac system in semiclassical gravity we normal order the stress-energy-momentum tensor.  The Einstein field equations we make use of are then
\begin{equation} \label{:EFE:}
G_{\mu\nu} = 8\pi G \langle {:\mathrel{\widehat{T}_{\mu\nu}}:} \rangle,
\end{equation}
which leads to sensible finite results.  While it may be beneficial to make use of a more sophisticated renormalization scheme, we find it useful to compare the system of equations derived from quantization and normal ordering with the multifield system of equations derived in the literature.

The only term in Eq.~(\ref{That b d}) that changes upon normal ordering is ${:\mathrel{\hat{d}_I \hat{d}^\dag_J}:} = -\hat{d}^\dag_J \hat{d}_I$ and 
\begin{equation} \label{:That b d:}
\begin{split}
\langle {:\mathrel{\widehat{T}_{\mu\nu}}:}\rangle  &= \sum_{I,J} \Bigl[
T_{\mu\nu}(f_I^+, f_J^+) \langle \hat{b}^\dag_I \hat{b}_J  \rangle
+ T_{\mu\nu}(f_I^+, f_J^-) \langle\hat{b}^\dag_I \hat{d}^\dag_J \rangle
\\
&\qquad
+ T_{\mu\nu}(f_I^-, f_J^+) \langle\hat{d}_I \hat{b}_J \rangle
- T_{\mu\nu}(f_I^-, f_J^-) \langle\hat{d}^\dag_J \hat{d}_I\rangle
\Bigr].
\end{split}
\end{equation}
The Einstein field equations in (\ref{:EFE:}) lead to the following two equations for determining the metric functions $\alpha(r)$ and $a(r)$ \cite{AlcubierreBook}:
\begin{equation} \label{metric field equations}
\begin{split}
\partial_r \alpha &= + \frac{\alpha(a^2-1)}{2r}
+ 4\pi G r \alpha a^2  \langle S\indices{^r_r} \rangle
\\
\partial_r a &= 
- \frac{a(a^2 - 1)}{2r}
+ 4\pi G r a^3 \langle \rho \rangle,
\end{split}
\end{equation}
where 
\begin{equation} \label{rho Srr}
\begin{split}
\langle \rho \rangle = \frac{1}{\alpha^2} \langle {:\mathrel{\widehat{T}_{tt}}:}  \rangle
\qquad
\langle S\indices{^r_r} \rangle = \frac{1}{a^2} \langle {:\mathrel{\widehat{T}_{rr}}:}  \rangle
\end{split}
\end{equation}
are the expectation values of the energy density and spatial stress.  The two equations in (\ref{metric field equations}), in combination with the radial equations of motion in (\ref{P radial eom}) and the normalization requirement in (\ref{norm req}), constitute the full system of equations to be solved for the static spherically symmetric semiclassical Einstein-Dirac system.  In the next section, we describe our methods for solving them numerically and present example solutions.  In the remainder of this section, we compute the expectation value $\langle {:\mathrel{\widehat{T}_{\mu\nu}}:} \rangle$, present the spherically symmetric equations for $T_{\mu\nu}(f_I^\pm, f_I^\pm)$, and construct $\langle \rho \rangle$ and $\langle S\indices{^r_r} \rangle$.

We compute expectation values using the basis states constructed in Sec.~\ref{sec:states}.  Physically, the basis states are states for a definite number of particles.  We do not consider the more general possibility of expectation values with states made from linear combinations of the basis states, as given in Eq.~(\ref{general state}), though it would be interesting to do so.  Using an arbitrary basis state from Eq.~(\ref{basis state}), the expectation values needed in Eq.~(\ref{:That b d:}) are
\begin{equation}
\begin{aligned}
\langle \hat{b}^\dag_I \hat{b}_J\rangle &= \delta_{IJ} N^b_I,&
\quad
\langle \hat{d}^\dag_J\hat{d}_I\rangle &= \delta_{IJ} N^d_I,
\\
\langle \hat{b}^\dag_I \hat{d}^\dag_J \rangle &= 0,&
\langle \hat{d}_I \hat{b}_J  \rangle &= 0.
\end{aligned}
\end{equation}
Plugging these into (\ref{:That b d:}), we have
\begin{equation} \label{<T> 0}
\langle {:\mathrel{\widehat{T}_{\mu\nu}}:} \rangle 
= \sum_I \left[
N_I^b T_{\mu\nu}(f_I^+, f_I^+)
- N_I^d T_{\mu\nu}(f_I^-, f_I^-) \right].
\end{equation}
Since our spacetime is spherically symmetric, we must only make use of spherically symmetric basis states.  As explained in Sec.~\ref{sec:states}, spherically symmetric states have all possible values of $m_j$ excited for each excited $j$.  We have then
\begin{equation} \label{<T>}
\begin{split}
\langle {:\mathrel{\widehat{T}_{\mu\nu}}:} \rangle 
&= \sum_{n,j} N_{nj}^b \sum_{m_j = -j}^j
T_{\mu\nu}(f_{njm_j+}, f_{njm_j+})
\\
&\qquad
- \sum_{n,j} N_{nj}^d \sum_{m_j = -j}^j
T_{\mu\nu}(f_{njm_j-}, f_{njm_j-}),
\end{split}
\end{equation}
where the mode functions $f_{njm_j\pm}$ are given in (\ref{mode functions}) and where $N_{nj}^b = 1$ for particles with excited quantum numbers $n$ and $j$ and equals 0 otherwise and similarly for $N_{nj}^d$ for antiparticles.

The final thing we need are the components of $T_{\mu\nu}(f_I^\pm, f_I^\pm)$ summed over $m_j$.  For the static spherically symmetric metric in (\ref{metric}), the stress-energy-momentum tensor must be diagonal and the nonvanishing components work out to
\begin{align}
\sum_{m_j = -j}^j
T_{tt}(f_I^\pm, f_I^\pm)
&=
\pm \frac{\omega_{nj}(2j+1)}{4\pi r^2}
\left(
|P_{nj+}|^2 
+ 
|P_{nj-}|^2 
\right)
\notag \\
\sum_{m_j = -j}^j
T_{rr}(f_I^\pm,f_I^\pm)
&=
\pm
\frac{a^2(2j+1)}{4\pi r^2 \alpha^2}
\biggl[ 
(\omega_{nj} - \alpha m_\psi)|P_{nj+}|^2
\notag \\
&\qquad
+ (\omega_{nj} + \alpha m_\psi)|P_{nj-}|^2
\notag \\[7pt]
&\qquad
- \frac{\alpha(2j+1)}{r} 
\text{Re}
\left( P_{nj+} P_{nj-}\right)
\biggr]
\notag \\
\sum_{m_j = -j}^j
T_{\theta\theta}(f_I^\pm,f_I^\pm)
&=
\pm \frac{(2j+1)^2}{8\pi  r \alpha} \text{Re}
\left( P_{nj+} P_{nj-} \right)
\notag \\
\sum_{m_j = -j}^j
T_{\phi\phi}(f_I^\pm,f_I^\pm)
&=
\pm \frac{(2j+1)^2}{8\pi r \alpha} \text{Re}
\left( P_{nj+} P_{nj-} \right) \sin^2\theta.
\label{Tff}
\end{align}
We explain how to derive these formulas in Appendix \ref{EM components}.  For completeness, we also present the off-diagonal component
\begin{equation}
\sum_{m_j = -j}^j
T_{tr}(f_I^\pm,f_I^\pm)
=
\mp \frac{\omega_{nj} a(2j+1)}{2\pi\alpha r^2} 
\text{Im} (P_{nj+} P_{nj-}).
\end{equation}
It is clear that this component will vanish, as required, if $P_{nj+}$ and $P_{nj-}$ are both purely real, which we will assume from this point forward.  We anticipated in Sec.~\ref{sec:eom sol} that $P_{nj\pm}$ would be purely real when we were able to write the radial equations of motion in Eq.~(\ref{P radial eom}) such that all $i$s canceled out.  We see now that $P_{nj\pm}$ being purely real leads to a spherically symmetric and static spacetime.

In the next section, we present example solutions.  For simplicity, we will consider $b$-type excitations only so that $N_I^d = 0$.  Combining Eqs.~(\ref{<T>}) and (\ref{Tff}), the energy density and spatial stress in (\ref{rho Srr}) are given by
\begin{align}
\langle \rho \rangle &= \sum_{n,j} N_{nj}^b
\frac{2j+1}{4\pi r^2 \alpha^2}
\omega_{nj}
\left(
P_{nj+}^2 
+ 
P_{nj-}^2 
\right)
\notag \\
\langle S\indices{^r_r} \rangle &= \sum_{n,j} N_{nj}^b
\frac{2j+1}{4\pi r^2 \alpha^2}
\biggl[ 
(\omega_{nj} - \alpha m_\psi)P_{nj+}^2
\notag \\
&\qquad
+ (\omega_{nj} + \alpha m_\psi)P_{nj-}^2
\notag \\[7pt]
&\qquad
- \frac{\alpha(2j+1)}{r} 
P_{nj+} P_{nj-}
\biggr].
\label{rho Srr full}
\end{align}


\section{Einstein-Dirac system in semiclassical gravity}
\label{ED SG}

Configurations of the semiclassical Einstein-Dirac system are described by self-consistent solutions to the equations of motion and the Einstein field equations.  In general, self-consistently solving this system of equations is highly nontrivial \cite{Diez-Tejedor:2011plw, Alcubierre:2022rgp}.  However, for the static spacetime we are considering, it is possible to classify solutions to the equations of motion as positive and negative frequency, as we saw in Sec.~\ref{sec:eom sol}.  This allows for the identification of a preferred vacuum state.  By using the basis states to the Hilbert space that are formed from excitations of the vacuum state we were able to compute the expectation value of the stress-energy-momentum tensor before actually having solved for the stress-energy-momentum tensor, as shown in Eq.~(\ref{<T> 0}).  These facts allow us to self-consistently solve the system of equations.  In Sec.~\ref{sec:numerical}, we write our system of equations in a form that is better suited for solving numerically and discuss our numerical methods.  In Sec.~\ref{sec:examples}, we present example solutions.


\subsection{Numerical methods}
\label{sec:numerical}

Our system of equations comprises of the radial equations of motion in (\ref{P radial eom}), the metric field equations in (\ref{metric field equations}), the expectation values of the energy density and spatial stress in (\ref{rho Srr full}), and the normalization requirement in (\ref{norm req}).  Since the metric in Eq.~(\ref{metric}) is written in terms of the Schwarzschild radial coordinate, we have also the mass function
\begin{equation}
m(r) \equiv \frac{r}{2G} \left[1 - \frac{1}{a^2(r)} \right],
\end{equation}
which gives the Arnowitt-Deser-Misner mass $M$ in the large $r$ limit, $m(r\rightarrow \infty) = M$.

These equations have scaling symmetries.  We can use these scaling symmetries to write the equations in terms of dimensionless variables, which is convenient to do when solving equations numerically.  We define the dimensionless variables
\begin{equation} \label{dim var}
\begin{aligned}
\bar{r} &\equiv m_\psi r,&
\qquad
\bar{\omega}_{nj} &\equiv \frac{\omega_{nj}}{m_\psi}
\\
\overline{P}_{nj\pm}(r) &= \sqrt{Gm_\psi} \, P_{nj\pm}(r),&
\qquad
\bar{m}(r) &\equiv Gm_\psi  m(r),
\end{aligned}
\end{equation}
where we note that $\alpha$, $a$, and $\mathcal{N}$ are already dimensionless.  When our system of equations is written in terms of these variables, all factors of $m_\psi$ and $G$ cancel out and they do not have to be specified.  

Although $\mathcal{N}$ is dimensionless, it is modified when written in terms of the dimensionless variables, becoming $\mathcal{N}_{nj} = (m_\text{Pl}/m_\psi)^2\overline{\mathcal{N}}_{nj}$, where $m_\text{Pl} = 1/\sqrt{G}$ is the Planck mass and
\begin{equation} \label{N bar}
\overline{\mathcal{N}}_{nj} \equiv 
\int_0^\infty d\bar{r} \, \frac{a}{\alpha}
\left( 
\overline{P}_{nj+}^2
+
\overline{P}_{nj-}^2
\right).
\end{equation}
From the normalization requirement in Eq.~(\ref{norm req}), $\mathcal{N}_{nj} = 1$ and the normalization requirement becomes
\begin{equation} \label{mass parameter}
\frac{m_\psi}{m_\text{Pl}} = \sqrt{\overline{\mathcal{N}}_{nj}}.
\end{equation}
In particular, all $\overline{\mathcal{N}}_{nj}$ must equal the same value.  This can be a nontrivial constraint on solutions.  We explain below how we impose this constraint.

Since the system of equations is independent of $m_\psi$ after moving to the dimensionless variables, one might expect solutions to be valid for arbitrary values of $m_\psi$, but this is not true.  Each solution is only valid for a single value of $m_\psi$ as given by (\ref{mass parameter}), since only for this specific value are the mode functions normalized.  Since $m_\psi$ is used to define the dimensionless variables, Eq.~(\ref{mass parameter}) leads to the additional results
\begin{equation}
t_\text{Pl} \omega_{nj} = \sqrt{\overline{\mathcal{N}}_{nj}} \, \bar{\omega}_{nj},
\qquad
\frac{M}{m_\text{Pl}} = \frac{\overline{M}}{\sqrt{\overline{\mathcal{N}}_{nj}}},
\end{equation}
where $t_\text{Pl} = \sqrt{G}$ is the Planck time and $\overline{M} = \bar{m}(\bar{r}\rightarrow \infty)$.

The radial equations of motion in (\ref{P radial eom}) and the metric field equations in (\ref{metric field equations}) are first order ordinary differential equations.  They may be solved by numerically integrating them outward from $r = 0$.  This requires inner boundary values for $\alpha$, $a$, and $\overline{P}_{nj\pm}$ and specification of the $\bar{\omega}_{nj}$.  

For $\alpha$, we use that the system of equations has an additional scaling symmetry, allowing us to define
\begin{equation}
\widetilde{\alpha}(\bar{r}) \equiv \frac{\alpha(\bar{r})}{\alpha(0)},
\quad
\widetilde{\omega}_{nj} \equiv \frac{\bar{\omega}_{nj}}{\alpha(0)},
\quad
\widetilde{P}_{nj\pm}(\bar{r}) \equiv \frac{\overline{P}_{nj\pm}(\bar{r})}{\sqrt{\alpha(0)}}.
\end{equation}
When the system of equations are written in terms of these quantities, $\alpha(0)$ cancels out and we have the inner boundary condition $\widetilde{\alpha}(0) = 1$.  We assume the spacetime is asymptotically Schwarzschild.  As a consequence, in the large $r$ limit $\alpha(r) = 1/a(r)$.  After a solution is found, $\alpha(0)$ can be determined from
\begin{equation}
\alpha(0) = \frac{1}{\widetilde{\alpha}(\bar{r}\rightarrow \infty)a(\bar{r}\rightarrow \infty)},
\end{equation}
which can then be used to give $\alpha(\bar{r}) = \alpha(0) \widetilde{\alpha}(\bar{r})$,  $\bar{\omega}_{nj} = \alpha(0) \widetilde{\omega}_{nj}$, and $\overline{P}_{nj\pm}(\bar{r}) = \sqrt{\alpha(0)} \widetilde{P}_{nj\pm}(\bar{r})$.

For the remaining fields, inner boundary values can be determined by plugging Taylor series expansions of $\widetilde{\alpha}(\bar{r})$, $a(\bar{r})$, and $\widetilde{P}_{nj\pm}(\bar{r})$ into the system of equations and then solving them for small $\bar{r}$.  One finds
\begin{equation}
\begin{split}
\widetilde{\alpha}(\bar{r}) &= 1 + O(\bar{r}^{2j_\text{min}+1})
\\
a(\bar{r}) &= 1 + O(\bar{r}^{2j_\text{min}+1})
\\
\widetilde{P}_{nj+}(\bar{r}) &= p_{nj} \bar{r}^{j+1/2} + O(\bar{r}^{j+5/2})
\\
\widetilde{P}_{nj-}(\bar{r}) &= p_{nj}\frac{\widetilde{\omega}_{nj} - 1}{2(j+1)} \bar{r}^{j+3/2} + O(\bar{r}^{j+7/2}),
\end{split}
\end{equation}
where $j_\text{min}$ is the smallest value of $j$ that is excited and where the $p_{nj}$ are undetermined constants.  

To solve the equations we still need values for the $\widetilde{\omega}_{nj}$ and the $p_{nj}$.  We choose to specify one of the $p_{nj}$ ourselves and to solve for the remaining $p_{nj}$ and for all of the $\widetilde{\omega}_{nj}$ using the shooting method.  That is, we begin with trial values for the remaining $p_{nj}$ and for all of the $\widetilde{\omega}_{nj}$.  We then vary these values until the normalization condition in (\ref{mass parameter}) and the outer boundary conditions are satisfied.  The normalization condition in (\ref{mass parameter}) requires that all $\overline{\mathcal{N}}_{nj}$ equal one another and the outer boundary conditions are that the expectation value of the energy density $\langle \rho \rangle$ asymptotically heads to zero.  This is achieved by requiring
\begin{equation}
\widetilde{P}_{nj\pm}(\bar{r}\rightarrow \infty) \rightarrow 0.
\end{equation}

The complete set of scaled equations that we solve numerically are
\begin{equation}
\begin{split}
\partial_{\bar{r}} \widetilde{P}_{nj\pm} 
&= \mp \frac{a}{\widetilde{\alpha}}
\left[ \left(\widetilde{\omega}_{nj} \pm \widetilde{\alpha}\right) \widetilde{P}_{nj\mp} - \frac{\widetilde{\alpha}}{\bar{r}} (j + 1/2) \widetilde{P}_{nj\pm} \right]
\\
\partial_{\bar{r}} \widetilde{\alpha} &= 
+ \frac{\widetilde{\alpha}(a^2-1)}{2\bar{r}}
+ 4\pi \bar{r} \widetilde{\alpha} a^2 \langle \overline{S}\indices{^r_r} \rangle
\\
\partial_{\bar{r}} a &= 
- \frac{a(a^2 - 1)}{2\bar{r}}
+ 4\pi \bar{r} a^3 \langle \bar{\rho} \rangle,
\end{split}
\end{equation}
where
\begin{equation}
\begin{split}
\langle \bar{\rho} \rangle &= \sum_{n,j} N_{nj}^b
\frac{2j+1}{4\pi \bar{r}^2 \widetilde{\alpha}^2}
\widetilde{\omega}_{nj}
\left(
\widetilde{P}_{nj+}^2 
+ 
\widetilde{P}_{nj-}^2 
\right)
\\
\langle \overline{S}\indices{^r_r} \rangle &= \sum_{n,j} N_{nj}^b
\frac{2j+1}{4\pi \bar{r}^2 \widetilde{\alpha}^2}
\biggl[ 
(\widetilde{\omega}_{nj} - \widetilde{\alpha}) \widetilde{P}_{nj+}^2
\\
&\qquad
+ (\bar{\omega}_{nj} + \widetilde{\alpha}) \widetilde{P}_{nj-}^2
\\[7pt]
&\qquad
- \frac{\widetilde{\alpha}(2j+1)}{\bar{r}} 
\widetilde{P}_{nj+} \widetilde{P}_{nj-}
\biggr],
\end{split}
\end{equation} 
along with
\begin{equation}
\begin{split}
\overline{\mathcal{N}} 
&=
\int_0^\infty d\bar{r} \, \frac{a}{\widetilde{\alpha}}
\left( 
\widetilde{P}_{nj+}^2
+
\widetilde{P}_{nj-}^2
\right)
\\
\bar{m} &= \frac{\bar{r}}{2}\left(1-\frac{1}{a^2}\right).
\end{split}
\end{equation}
Solutions to these equations are distinguished by the quantum numbers $n$ and $j$ and the value of one of the $p_{nj}$.  If $N_{nj}^b \neq 0$ for a single value of $n$ and for $j=1/2$, the equations above are equivalent to those in \cite{Finster:1998ws}.  If $N_{nj}^b \neq 0$ for a single value of $n$ and a single value of $j$, the equations are consistent with those in \cite{Finster:1998ju}, which studied the Einstein-Dirac-Maxwell system for arbitrary $j$.  Our framework therefore reproduces the results in the literature.  However, all previous studies of the spherically symmetric Einstein-Dirac system, including Refs.~\cite{Finster:1998ws, Finster:1998ju}, postulated $2j+1$ independent Dirac fields.  This was explained as being necessary to preserve spherical symmetry, where a different Dirac field was needed for each value of $m_j$.  In our framework, we have a single quantum Dirac field.  We preserve spherical symmetry by exciting all values of $m_j$, so that our system is a self-gravitating configuration of $2j+1$ identical quantum particles.  Further, our framework can straightforwardly accommodate multiple values of $n$ and $j$, which has not previously been considered.


\subsection{Example solutions}
\label{sec:examples}

\begin{figure*}
\centering
\includegraphics[width=7in]{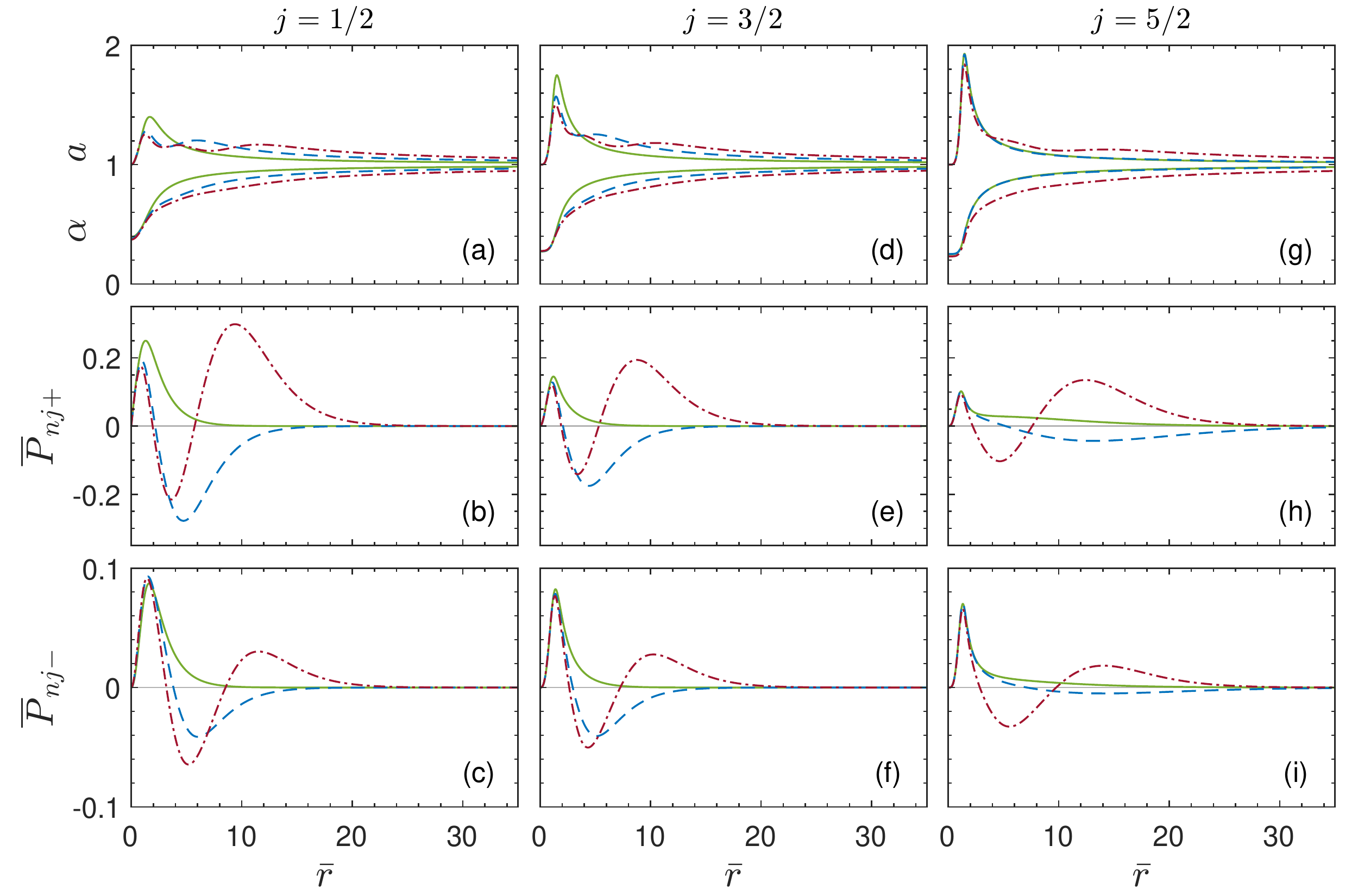}
\caption{Example semiclassical configurations with a single value of $n$ and a single value of $j$ excited.  The left column (a--c) is for  $j=1/2$, the middle column (d--f) is for $j=3/2$, and the right column (g--i) is for $j=5/2$.  In each plot, the solid green curve is for $n = 0$, the dashed blue curve is for $n=1$, and the dot-dashed maroon curve is for $n=2$.}
\label{fig:1}
\end{figure*}

\begin{figure*}
\centering
\includegraphics[width=7in]{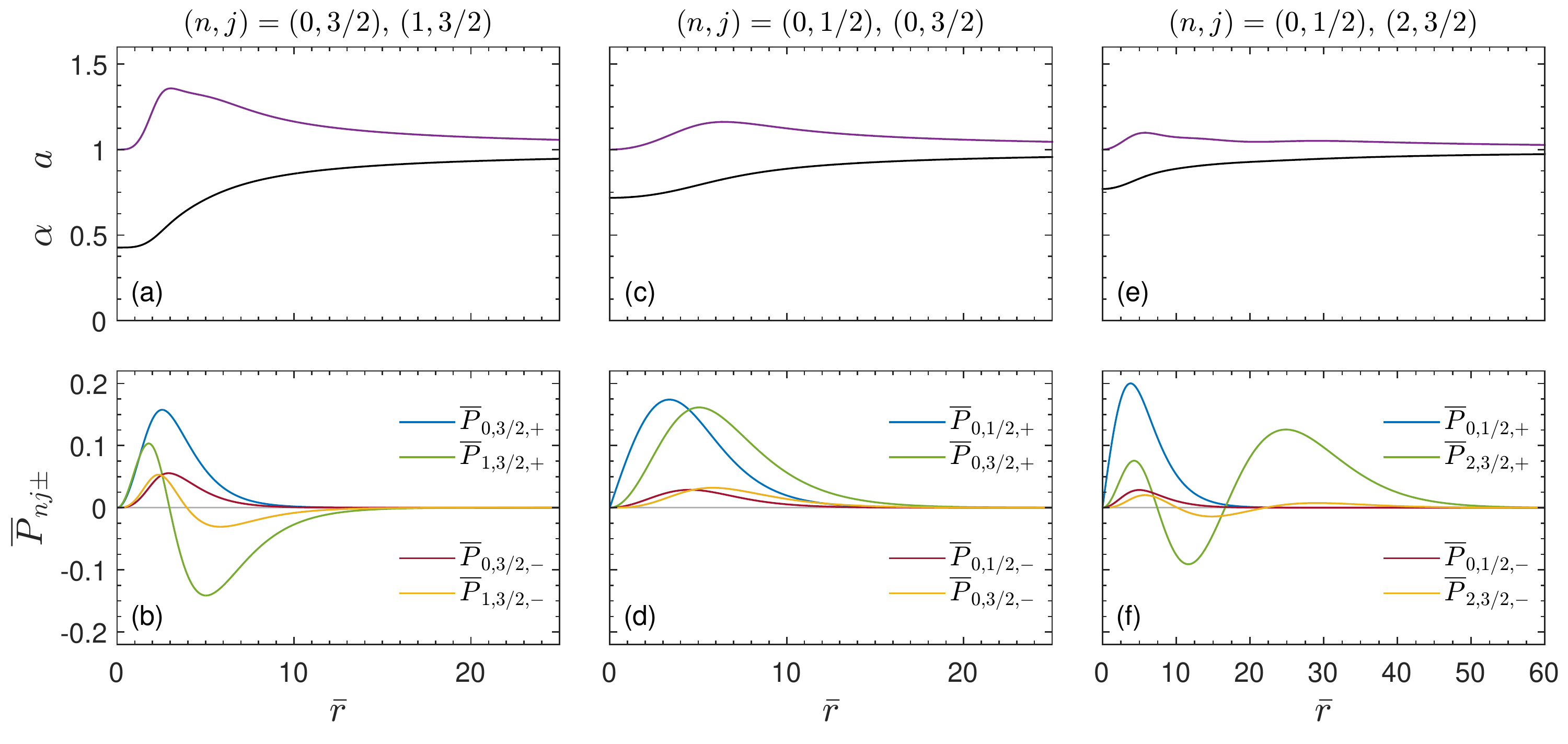}
\caption{Example semiclassical configurations with two pairs of $(n,j)$ excited, as indicated above each column.  The left column (a, b) displays a multi-$n$ configuration, the middle column (c, d) displays a multi-$j$ configuration, and the right column (e, f) displays a multi-$n$ multi-$j$ configuration. }
\label{fig:2}
\end{figure*}

\setlength{\tabcolsep}{17pt}
\begin{table*} 
\normalsize
\begin{tabular}{cccccccc}
$n,\, j$ & $p_{nj}$  &  $\widetilde{\omega}_{nj}$ & $\alpha(0)$ & $\overline{\mathcal{N}}_{nj}$ & $t_\text{Pl} \omega_{nj}$ &  $m_\psi/m_\text{Pl}$ & $M/m_\text{Pl}$ 
\\
\hline\hline
$0,\, 1/2$ & 0.5 & 1.8754 & 0.3933 & 0.2931 & 0.3993 & 0.5414 & 1.0976
\\ \hline
$0,\, 3/2$ & 0.5 & 2.8789 & 0.2729 & 0.1231 & 0.2756 & 0.3508 & 1.8786
\\\hline
$0,\, 5/2$ & 0.5 & 3.8272 & 0.2501 & 0.0693 & 0.2520 & 0.2633 & 2.6760
\\\hline
$1,\, 1/2$ & 0.5 & 2.1492 & 0.3815 & 0.5636 & 0.6155 & 0.7507 & 1.5210 
\\\hline
$1,\, 3/2$ & 0.5 & 2.9843 & 0.2750 & 0.2630 & 0.4208 & 0.5128 & 2.3168
\\\hline
$1,\, 5/2$ & 0.5 & 3.8289 & 0.2525 & 0.0842 & 0.2805 & 0.2902 & 2.7308
\\\hline
$2,\, 1/2$ & 0.5 & 2.2644 & 0.3726 & 0.8869 & 0.7946 & 0.9418 & 1.8901
\\\hline
$2,\, 3/2$ & 0.5 & 3.0415 & 0.2776 & 0.4036 & 0.5365 & 0.6353 & 2.7368  
\\\hline
$2,\, 5/2$ & 0.5 & 3.8628 & 0.2329 & 0.2566 & 0.4558 & 0.5066 & 3.5119  
\\ \hline
$0,\, 3/2$ & 0.1    & 1.6404 & 0.4267 & 0.1613 & 0.2811 & 0.4017 & 3.2694 
\\
$1,\, 3/2$ & 0.1184 & 1.8978 &        & 0.1613 & 0.3252 &        & 
\\ \hline
$0,\, 1/2$ & 0.1    & 1.1314 & 0.7176 & 0.1846 & 0.3488 & 0.4296 & 2.4796
\\
$0,\, 3/2$ & 0.0206 & 1.2127 &        & 0.1846 & 0.3738 &        & 
\\ \hline
$0,\, 1/2$ & 0.1    & 1.1123 & 0.7692 & 0.2688 & 0.4436 & 0.5185 & 3.0401
\\
$2,\, 3/2$ & 0.0116 & 1.2346 &        & 0.2688 & 0.4923 &        & 
\end{tabular}
\caption{Values of various quantities for all example configurations shown in Figs.~\ref{fig:1} and \ref{fig:2}.}
\label{table}
\end{table*}

In this subsection, we present some static spherically symmetric semiclassical Einstein-Dirac configurations.  At the end of Sec.~\ref{sec:semiclassical gravity}, we explained that we are focusing on $b$-type excitations only and are therefore not including antiparticles.  Configurations are labeled by the quantum numbers $n$ and $j$, where $j=1/2, 3/2, 5/2, \ldots$, and the value of one of the $p_{nj}$.

We begin with configurations that have a single $n$ and a single $j$ excited.  In this case, the system of equations is relatively easy to solve because there is only a single shooting parameter, $\widetilde{\omega}_{nj}$.  In Fig.~\ref{fig:1}, we display various solutions.  The left column is for $j = 1/2$, the middle column is for $j = 3/2$, and the right column is for $j=5/2$.  In each plot, there is a curve for $n=0$, 1, and 2.  One thing to take note of is that we can see that the quantum number $n$ is equal to the number of nodes in $\overline{P}_{nj\pm}$ and can therefore be interpreted as a radial quantum number.  The values of various quantities for these solutions are listed in Table \ref{table}.

We now present multi-$n$ and multi-$j$ solutions, which has not previously been considered in the literature.  Specifically, we show solutions with two $(n,j)$ pairs excited.  Since each pair has associated with it a $p_{nj}$, one of which we specify, and an $\widetilde{\omega}_{nj}$, we have three shooting parameters.  In Fig.~\ref{fig:2}, we show in the left column a multi-$n$ solution, in the middle column a multi-$j$ solution, and in the right column a multi-$n$ multi-$j$ solution.  The values of various quantities for these solutions are listed in Table \ref{table}.

One question we might ask about these solutions is whether or not they are stable.  The nonlinear stability of the $n=0, \, j=1/2$ solutions was studied in \cite{Daka:2019iix} for classical gravity.  In that work, the normalization requirement in (\ref{norm req}) was not used.  However, for single-$n$, single-$j$ solutions, the normalization requirement is trivially satisfied for a particular value of the mass parameter, as given by (\ref{mass parameter}).  It was shown in \cite{Daka:2019iix} that a large class of $n=0, \, j=1/2$ solutions are nonlinearly stable and that the unstable solutions migrate to stable solutions.  The nonlinear stability of higher $n$ and higher $j$ solutions has not been studied.  Unfortunately, it is unclear how to perform a nonlinear stability analysis in quantum field theory with the semiclassical gravity approximation, since in this case the metric is no longer static.

We end this section with a brief comparison of the solutions presented here for spin-1/2 fermions and those presented in \cite{Alcubierre:2022rgp} for spin-0 bosons.  At first glance, our Fig.~\ref{fig:2} and Fig.~1 in \cite{Alcubierre:2022rgp} suggest a similar structure for radial functions and metric fields.  Unsurprisingly, there exists a significant difference between the systems because fermions obey the Pauli exclusion principle, which is implemented through the normalization requirement in (\ref{norm req}).  The analogous expression in \cite{Alcubierre:2022rgp} gives the number of bosons in each state and is not required to be equal to 1.  In practice, this makes the fermionic system more challenging to solve for multi-$n$, multi-$j$ solutions, since in this case the normalization requirement is a nontrivial constraint on solutions.  For single-$n$, single-$j$ solutions, the fermionic system is only valid for a single value of the mass parameter, as given by (\ref{mass parameter}), while the bosonic system in \cite{Alcubierre:2022rgp} is valid for arbitrary values of the analogous mass parameter.


\section{Conclusion}
\label{sec:conclusion}

We constructed static spherically symmetric self-gravitating configurations of quantum spin-1/2 particles in quantum field theory using the semiclassical gravity approximation.  We began with a single Dirac field minimally coupled to general relativity.  We canonically quantized the Dirac field in a static spherically symmetric curved spacetime background.  By considering a static spacetime, we were able to identify a preferred vacuum state, from which we constructed spherically symmetric basis states to the Hilbert space through repeated application of creation operators.  Using these states, we were able to compute the expectation value of the stress-energy-momentum tensor.  This allowed us to derive a system of equations whose solutions describe static spherically symmetric semiclassical Einstein-Dirac configurations.  We self-consistently solved these equations and presented example configurations.

Limiting cases of the semiclassical system of equations that we derived agree with the multifield system of equations derived in the literature.  As a consequence, the multifield configurations given in, say, Ref.~\cite{Finster:1998ws} are equivalent to the analogous semiclassical configurations, at least for the normal ordering renormalization scheme used in this work.  Although equivalent, the semiclassical equations are derived from the excitations of a single quantum Dirac field, while the multifield equations require introducing multiple independent Dirac fields.  The semiclassical system of equations are also more general, since they can naturally accommodate multi-$n$ and multi-$j$ configurations, examples of which we presented, as well as configurations with both particles and antiparticles.

\appendix

\section{Hermiticity of $\widehat{H}$}
\label{sec:hermiticity}

In this appendix, we show that the operator $\widehat{H}$ in (\ref{Hhat def}) is Hermitian with respect to the inner product
\begin{equation} 
\begin{split}
( u_{I}, u_{J} )
&= \int_\Sigma d^3x \sqrt{\det(\gamma_{ij})}\, u_I^\dag u_J
\\
&= \int_{\Sigma} dr d\theta d\phi \, a r^2 \sin\theta \, u_{I}^\dag  u_{J},
\end{split}
\end{equation}
that is, we show that
\begin{equation} \label{hermitian H}
( u_{I}, \widehat{H} u_{J} )
= ( \widehat{H} u_{I}, u_{J} ).
\end{equation}

We begin with some convenient results.  From (\ref{Dirac rep}), we have $(\tilde{\gamma}^0)^\dag = -\tilde{\gamma}^0$, $(i\tilde{\gamma}^0)^\dag = i\tilde{\gamma}^0$, and $(\tilde{\gamma}^j)^\dag = \tilde{\gamma}^j$.  It follows from (\ref{gamma t r t p}) that $(\tilde{\gamma}^r)^\dag = \tilde{\gamma}^r$.  From (\ref{Sigma j}), we can see that $\tilde{\gamma}^0$ commutes with $\widehat{\Sigma}_j$.  It is obvious that $\tilde{\gamma}^0$ commutes with $\widehat{\mathbf{L}}$, and hence $\tilde{\gamma}^0$ commutes with $\widehat{\mathbf{J}}$ and $\widehat{J}^{\,2}$.  Since $\tilde{\gamma}^0$ commutes with all terms in $\widehat{K}$ in (\ref{K J L}), $\tilde{\gamma}^0$ commutes with $\widehat{K}$.  Since $\tilde{\gamma}^0$ also commutes with all terms inside the parentheses in (\ref{K J L}), we have $\widehat{K}^\dag = \widehat{K}$, where we used the well-known identities $\widehat{J}^{\,2\dag} = \widehat{J}^{\, 2}$ and $\widehat{L}^{2\dag} = \widehat{L}^2$.

Having established these results, we turn to $\widehat{H}$ in (\ref{Hhat def}),
\begin{equation} 
\widehat{H} = \frac{i\alpha}{a} \tilde{\gamma}^0 \tilde{\gamma}^r
\left(\partial_r  + \frac{\partial_r \alpha}{2\alpha}  + \frac{1}{r} \right)
- \frac{\alpha}{r} \tilde{\gamma}^r \widehat{K}
- i\alpha m_\psi \tilde{\gamma}^0.
\end{equation}
We have previously mentioned that $\tilde{\gamma}^r$ commutes with $\widehat{K}$ and thus, for each of these terms,
\begin{equation} \label{dag results}
\begin{split}
(i\tilde{\gamma}^0 \tilde{\gamma}^r)^\dag = -i  \tilde{\gamma}^0 \tilde{\gamma}^r,
\quad
(\tilde{\gamma}^r \widehat{K})^\dag = \tilde{\gamma}^r \widehat{K},
\quad
(i\tilde{\gamma}^0)^\dag = i \tilde{\gamma}^0.
\end{split}
\end{equation}
Inserting $\widehat{H}$ into the inner product gives
\begin{equation}
( u_{I}, \widehat{H} u_{J} ) = U_1 + U_2 + U_3,
\end{equation}
where
\begin{align}
U_1 &= \int_{\Sigma} dr d\theta d\phi \, a r^2 \sin\theta \, u_{I}^\dag  
\left[\frac{i\alpha}{a} \tilde{\gamma}^0 \tilde{\gamma}^r
\left(\partial_r  + \frac{\partial_r \alpha}{2\alpha}  + \frac{1}{r} \right)
u_{J} \right]
\notag \\
U_2 &= \int_{\Sigma} dr d\theta d\phi \, a r^2 \sin\theta \, u_{I}^\dag 
\left( - \frac{\alpha}{r} \tilde{\gamma}^r \widehat{K} u_J \right)
\notag \\
U_3 &= \int_{\Sigma} dr d\theta d\phi \, a r^2 \sin\theta \, u_{I}^\dag 
\left( - i\alpha m_\psi \tilde{\gamma}^0 u_J \right).
\end{align}
Using (\ref{dag results}), we immediately have
\begin{equation}
\begin{split}
U_2 &= \int_{\Sigma} dr d\theta d\phi \, a r^2 \sin\theta \, \left(  - \frac{\alpha}{a} \tilde{\gamma}^r \widehat{K} u_{I}\right)^\dag 
u_J
\\
U_3 &= \int_{\Sigma} dr d\theta d\phi \, a r^2 \sin\theta \, 
\left( - i\alpha m_\psi \tilde{\gamma}^0 u_{I} \right)^\dag 
u_J,
\end{split}
\end{equation}
and these terms in $\widehat{H}$ are Hermitian.

$U_1$ requires a little more work because of the $r$-derivative.  We have first, using (\ref{dag results}),
\begin{equation}
\begin{split}
U_1 &= \int_{\Sigma} dr d\theta d\phi \, \alpha r^2 \sin\theta \, 
\\
&\qquad \times
\left(-i\tilde{\gamma}^0 \tilde{\gamma}^r u_{I}\right)^\dag  
\left(\partial_r  + \frac{\partial_r \alpha}{2\alpha}  + \frac{1}{r} \right)
u_{J}.
\end{split}
\end{equation}
Some of the terms can be rewritten using
\begin{equation} \label{ibp formula}
\alpha r^2 \left(\partial_r + \frac{\partial_r \alpha}{2\alpha} + \frac{1}{r} \right)  u_J 
= \frac{1}{2} \partial_r (\alpha r^2 u_J)  
+ \frac{1}{2} \alpha r^2 \partial_r u_J.
\end{equation}
Performing an integration by parts and assuming the solutions decay sufficiently quickly at spatial infinity so that the boundary terms can be dropped, we have
\begin{align}
U_1 &= \int_{\Sigma} dr d\theta d\phi \, \sin\theta
\notag
\\
&\qquad \times
\frac{1}{2} \Bigl\{ \alpha r^2
\partial_r (i\tilde{\gamma}^0 \tilde{\gamma}^r  u_I)
+ \partial_r \left[ \alpha r^2 \left(i\tilde{\gamma}^0 \tilde{\gamma}^r  u_{I}\right) \right] \Bigr\}^\dag u_J
\notag \\
&= \int_{\Sigma} dr d\theta d\phi \, \alpha r^2\sin\theta
\notag
\\
&\qquad \times \left[\left( \partial_r + \frac{\partial_r \alpha}{2\alpha} + \frac{1}{r}\right) i \tilde{\gamma}^0 \tilde{\gamma}^r u_I \right]^\dag u_J
\notag \\
&= \int_{\Sigma} dr d\theta d\phi \, a r^2 \sin\theta \,
\notag \\
&\qquad \times
\left[\frac{i\alpha}{a} \tilde{\gamma}^0 \tilde{\gamma}^r
\left(\partial_r  + \frac{\partial_r \alpha}{2\alpha}  + \frac{1}{r} \right)
u_{I} \right]^\dag
u_{J},
\end{align}
where in the second equality we again used (\ref{ibp formula}).  We have now shown that $\widehat{H}$ satisfies the Hermiticity condition in Eq.~(\ref{hermitian H}).


\section{$\sum_{m_j = -j}^j T_{\mu\nu}(f_I^\pm, f_I^\pm)$}
\label{EM components}

In this appendix, we explain how the formulas in Eq.~(\ref{Tff}) are derived.  These formulas give the diagonal components of the spherically symmetric
\begin{equation}
\sum_{m_j = -j}^j
T_{\mu\nu}(f_{njm_j\pm}, f_{njm_j\pm}),
\end{equation} 
where $T_{\mu\nu}(f_I^\pm, f_I^\pm)$ is defined in Eq.~(\ref{T(fI, fJ)}) and is evaluated with the same mode function in both slots and where the mode functions are given in Eq.~(\ref{mode functions}).

It is possible to rewrite $T_{\mu\nu}(f_I^\pm,f_I^\pm)$ into the more convenient form
\begin{align} 
T_{\mu\nu}(f_I^\pm,f_I^\pm)
&= 
-\frac{1}{2}\text{Re} \left(
\bar{f}_I^\pm \gamma_\mu \partial_\nu f_I^\pm
+\bar{f}_I^\pm \gamma_\nu \partial_\mu f_I^\pm \right)
\notag \\
&\qquad
+ \frac{1}{4} \bar{f}_I^\pm \left( 
\{\gamma_\mu,  \Gamma_\nu\} +  \{\gamma_\nu,  \Gamma_\mu\}
\right) f _I^\pm,
\label{T ff rewrite}
\end{align}
where $\Gamma_\mu$ is the spinor connection in (\ref{spinor connection}).  Notice that the $\gamma$-matrices have lowered indices.  Since $\gamma_{\mu} = g_{\mu\nu} \gamma^\nu$, it follows from the metric in (\ref{metric}) and the vierbein in (\ref{vierbein}) that
\begin{equation}
\gamma_t = - \alpha \tilde{\gamma}^t,
\quad
\gamma_r = a \tilde{\gamma}^r,
\quad
\gamma_\theta = r \tilde{\gamma}^\theta,
\quad
\gamma_\phi = r\sin\theta\, \tilde{\gamma}^\phi.
\end{equation}
The only nonvanishing $\{\gamma_\mu,  \Gamma_\nu\} +  \{\gamma_\nu,  \Gamma_\mu\}$ can be shown to be $\{\gamma_t,  \Gamma_\theta\} + \{\gamma_\theta,  \Gamma_t\}$ and $\{\gamma_t,  \Gamma_\phi\} + \{\gamma_\phi,  \Gamma_t\}$ which contribute to the off-diagonal components of $T_{\mu\nu}(f_I^\pm,f_I^\pm)$ and will not be considered.

In the course of deriving the formulas in Eq.~(\ref{Tff}), we make heavy use of the sum formula
\begin{equation} \label{spin angle sum}
\sum_{m_j = -j}^j
(\mathcal{Y}_{j\pm1/2}^{m_j})^\dag \mathcal{Y}_{j\pm1/2}^{m_j}
= 
\frac{2j+1}{4\pi},
\end{equation}
where the $\mathcal{Y}_{j\pm1/2}^m$ are the spin-angle functions in (\ref{spin angle}).  This formula can be derived by changing the summation variable to $m_\ell$ and then using the standard sum formula for spherical harmonics, $\sum_{m_\ell = -\ell}^\ell |Y_\ell^{m_\ell}|^2 = (2\ell+1)/4\pi$.

The $tt$ component in Eq.~(\ref{Tff}) is obtained by straightforward calculations after plugging the mode functions in Eq.~(\ref{mode functions}) into (\ref{T ff rewrite}) and then summing over $m_j$.  For the $rr$ component, one must also use Eq.~(\ref{gamma r spinor}), but rewritten for the mode functions in the form in (\ref{mode functions}).  The result contains $r$-derivatives of $P_{nj\pm}$, which can be removed using the radial equations of motion in (\ref{P radial eom}).

The $\theta\theta$ and $\phi\phi$ components in Eq.~(\ref{Tff}) can be derived as follows.  For a spherically symmetric spacetime, it must be that $T_{\phi\phi} = T_{\theta\theta}\sin^2\theta$.  We can therefore write
\begin{equation}
T_{\theta\theta}(f_I^\pm, f_I^\pm) = \frac{1}{2} \left[
T_{\theta\theta}(f_I^\pm, f_I^\pm) \
+ \frac{T_{\phi\phi}(f_I^\pm, f_I^\pm)}{\sin^2\theta} \right].
\end{equation}
Using Eq.~(\ref{T ff rewrite}) for $T_{\theta\theta}$ and $T_{\phi\phi}$ inside the square brackets, this can be written as
\begin{align} \label{T theta theta}
T_{\theta\theta}(f_I^\pm, f_I^\pm)
&= - \frac{r}{2} \text{Re}\left[ \bar{f}_I^\pm \left( \tilde{\gamma}^\theta \partial_\theta + \frac{1}{\sin\theta} \tilde{\gamma}^\phi \partial_\phi \right) f_I^\pm \right]
\notag \\
&= - \frac{r}{2} \text{Re}\left[ \bar{f}_I^\pm 
\tilde{\gamma}^r \left( 1 + i \tilde{\gamma}^0 \widehat{K}\right) f_I^\pm \right],
\end{align}
where the operator $\widehat{K}$, first defined in (\ref{K def}), can be written
\begin{equation} 
\widehat{K} 
= -i\tilde{\gamma}^0\left( 1  -\tilde{\gamma}^r \tilde{\gamma}^\theta \partial_\theta
- \frac{1}{\sin\theta}\tilde{\gamma}^r \tilde{\gamma}^\phi \partial_\phi \right).
\end{equation}
Since $\widehat{K}f_I^\pm = \pm(j+1/2)f_I^\pm$, it is straightforward to evaluate (\ref{T theta theta}).  After summing over $m_j$, the $\theta\theta$ and $\phi\phi$ components in Eq.~(\ref{Tff}) can be obtained.


\section{Solutions to the classical equations of motion using the alternative choice for vierbein}
\label{sec:alt eom}

In Sec.~\ref{sec:eom sol}, we solved the classical equations of motion, i.e.~the Dirac equation,
\begin{equation} \label{app eom}
\left[\gamma^\mu (\partial_\mu - \Gamma_\mu) - m_\psi \right] f_I = 0,
\end{equation}
using the vierbein in (\ref{vierbein}).  In this appendix, we instead solve the Dirac equation using the alternative choice for the vierbein in (\ref{alt vierbein}).  This alternative choice leads to the angular dependence of the Dirac field being described by spin weighted spherical harmonics \cite{Newman:1966ub, Penrose:1986ca}, instead of by standard spherical harmonics and two-component spinors (for a pedagogical review of spin weighted spherical harmonics, see Appendix D of \cite{AlcubierreBook}).  We find the derivation presented in this appendix to be simpler than that presented in Sec.~\ref{sec:eom sol}.  However, a typical understanding of quantum angular momentum is based on spherical harmonics and two-component spinors and not on spin weighted spherical harmonics.  For this reason, we focused on the solutions derived in Sec.~\ref{sec:eom sol}.

For the vierbein in (\ref{alt vierbein}), the spinor connection works out to be
\begin{equation} \label{alt spinor connection}
\begin{split}
\Gamma_t &= \tilde{\gamma}^0 \tilde{\gamma}^3 \frac{\partial_r\alpha}{2a}
\\
\Gamma_r &= 0
\\
\Gamma_\theta &= \tilde{\gamma}^3 \tilde{\gamma}^2 \frac{1}{2 a} 
\\
\Gamma_\phi &= \tilde{\gamma}^3 \tilde{\gamma}^1 \frac{\sin\theta}{2 a}
+\tilde{\gamma}^2 \tilde{\gamma}^1 \frac{\cos\theta}{2}
,
\end{split}
\end{equation}
from which one can derive
\begin{equation} \label{alt gamma Gamma}
\begin{split}
\gamma^\mu \Gamma_\mu
&= 
- \tilde{\gamma}^3 \frac{1}{a}\left( \frac{\partial_r\alpha}{2\alpha} + 1 \right)
- \tilde{\gamma}^2 \frac{\cot\theta}{2 r},
\end{split}
\end{equation}
which is needed in (\ref{app eom}).

Just as in Sec.~\ref{sec:eom sol}, we drop the subscripted $I$ on $f_I$ and allow $f$ to have arbitrary dependencies, $f = f(t,r,\theta,\phi)$.  Using the vierbein in (\ref{alt vierbein}) and Eq.~(\ref{alt gamma Gamma}), the equations of motion can be written
\begin{equation} \label{H eom}
\widehat{H} f = i \partial_t f,
\end{equation}
where 
\begin{align}
\widehat{H} &\equiv \frac{i\alpha}{a} \left(\partial_r
+ \frac{\partial_r\alpha}{2\alpha} + \frac{1}{r}
\right) \tilde{\gamma}^0\tilde{\gamma}^3
- i\alpha m_\psi \tilde{\gamma}^0
+ \frac{\alpha}{r} \tilde{\gamma}^3 \widehat{\Omega}
\notag
\\
\widehat{\Omega} &\equiv i\left(\partial_\theta 
+ \frac{\cot\theta}{2}
\right) 
\tilde{\gamma}^3\tilde{\gamma}^0\tilde{\gamma}^2 
+ \frac{i}{\sin\theta} \partial_\phi 
\tilde{\gamma}^3\tilde{\gamma}^0\tilde{\gamma}^1.
\label{H Omega def}
\end{align} 
It is not difficult to show that the operators $i\partial_t$, $\widehat{H}$, and $\widehat{\Omega}$ commute with one another.  We can therefore assume $f$ is a simultaneous eigenfunction of all three operators.  

For $i\partial_t$ we have the eigenvalue equation $i\partial_t f = \omega f$.  One can show that $i\partial_t$ is Hermitian with respect to the inner product in Eq.\ (\ref{inner product 2}), $(f,i\partial_t f) = (i\partial_t f, f)$, and hence $\omega$ is real.  We now assume that the time dependence of $f$ is separable, so that $f$ can be written
\begin{equation} \label{f exp u}
f(t,r,\theta,\phi) = e^{-i\omega t} u(r,\theta,\phi).
\end{equation}
The remaining eigenvalue equations are then
\begin{equation} \label{H eom 2}
\widehat{H} u = \omega u,
\end{equation}
which is our equation of motion, and
\begin{equation} \label{angular eval eq}
\widehat{\Omega} u = \lambda u,
\end{equation}
where $\lambda$ are the eigenvalues of $\widehat{\Omega}$.

The next step in our solution is to assume that the radial ($r$) and angular ($\theta,\phi$) dependence separates,
\begin{equation}
u(r,\theta,\phi) = 
\begin{pmatrix}
R_1(r) \Theta_1(\theta,\phi) \\
R_2(r) \Theta_2(\theta,\phi) \\
R_3(r) \Theta_3(\theta,\phi) \\
R_4(r) \Theta_4(\theta,\phi) \\
\end{pmatrix}.
\end{equation}
We will find that Eq.~(\ref{H eom 2}) leads to the radial equations of motion and Eq.~(\ref{angular eval eq}) determines the angular dependence.

We focus first on the angular equations that follow from (\ref{angular eval eq}).  Separating (\ref{angular eval eq}) into four equations, we find
\begin{equation} \label{angular eom}
\begin{split}
\lambda R_1 \Theta_1 &= +iR_2\eth_+^{(-1/2)} \Theta_2
\\
\lambda R_2 \Theta_2 &= +iR_1\eth_-^{(+1/2)} \Theta_1
\\
\lambda R_3 \Theta_3 &= -iR_4\eth_+^{(-1/2)} \Theta_4
\\
\lambda R_4 \Theta_4 &= -iR_3\eth_-^{(+1/2)} \Theta_3,
\end{split}
\end{equation} 
where
\begin{equation}
\eth_\pm^{(s)}
= 
- \partial_\theta
\pm s \cot\theta
\mp \frac{i}{\sin\theta} \partial_\phi
\end{equation} 
are the raising and lowering operators for spin weighted spherical harmonics.  The effect these operators have on spin weighted spherical harmonics  ${_s Y_{j m_j}}$ is
\begin{equation} \label{raise lower Y eqs}
\eth^{(s)}_\pm  ({_s Y_{j m_j}}) = \pm \sqrt{(j \mp s)(j \pm s + 1)} ( {_{s\pm 1} Y_{j m_j}}),
\end{equation} 
where $s$ is the spin weight.  It follows from (\ref{raise lower Y eqs}) that spin weighted spherical harmonics satisfy the eigenvalue equations
\begin{equation} \label{spin weighted evalue eqs}
\begin{split}
\eth_-^{(s+1)} \eth_+^{(s)} {_sY_{j m_j}}
&= - [j(j+1)-s(s+1)] {_sY_{j m_j}}
\\
\eth_+^{(s-1)} \eth_-^{(s)} {_sY_{j m_j}}
&= - [j(j+1)-s(s-1)] {_sY_{j m_j}}.
\end{split}
\end{equation}
By applying the appropriate raising or lowering operator to each equation in (\ref{angular eom}) we can derive
\begin{equation} \label{spin weight eqs}
\begin{split}
\eth_-^{(+1/2)}\eth_+^{(-1/2)} \Theta_2
&=  -\lambda^2 \Theta_2
\\
\eth_+^{(-1/2)} \eth_-^{(+1/2)} \Theta_1
&= -\lambda^2 \Theta_1
\\
\eth_-^{(+1/2)} \eth_+^{(-1/2)} \Theta_4
&= -\lambda^2 \Theta_4
\\
\eth_+^{(-1/2)} \eth_-^{(+1/2)} \Theta_3
&= -\lambda^2 \Theta_3,
\end{split}
\end{equation}
where the radial parts cancel out and we obtain eigenvalue equations for spin weighted spherical harmonics.  As a consequence, $\Theta_1$ and $\Theta_3$ are spin weighted spherical harmonics with $s = +1/2$ and $\Theta_2$ and $\Theta_4$ are spin weighted spherical harmonics with $s=-1/2$.  $\Theta_1$ and $\Theta_3$ must then be proportional to one another and similarly for $\Theta_2$ and $\Theta_4$.  Proportionality constants can be absorbed into the radial part and we use the same convention used in \cite{Daka:2019iix},
\begin{equation} \label{T Y}
\Theta_1 = \Theta_3 = {_{(+1/2)} Y_{_j m_j}},
\qquad
\Theta_2 = -\Theta_4 = {_{(-1/2)} Y_{_j m_j}}.
\end{equation}
Plugging these into (\ref{spin weight eqs}) and comparing to (\ref{spin weighted evalue eqs}), we find the eigenvalues
\begin{equation} \label{lambda evalue}
\lambda = \mp (j+1/2).
\end{equation}
Plugging Eqs.~(\ref{T Y}) and (\ref{lambda evalue}) into Eq.~(\ref{angular eom}), the angular terms cancel and we find
\begin{equation} \label{R equiv}
R_2 = \pm i R_1, \qquad
R_4 = \pm iR_3,
\end{equation}
where the upper/lower signs in this equation are associated with those for the eigenvalues in (\ref{lambda evalue}).  The eigenspinors associated with the eigenvalues in (\ref{lambda evalue}) are then
\begin{equation}
u_\pm(r,\theta,\phi) = 
\begin{pmatrix}
\hphantom{\pm i} R_{1\pm}(r) {_{(+1/2)} Y_{_j m_j}} (\theta,\phi) \\ 
\pm iR_{1\pm}(r) {_{(-1/2)} Y_{_j m_j}} (\theta,\phi) \\ 
\hphantom{\pm i} R_{3\pm}(r) {_{(+1/2)} Y_{_j m_j}} (\theta,\phi)\\ 
\mp iR_{3\pm}(r) {_{(-1/2)} Y_{_j m_j}} (\theta,\phi)
\end{pmatrix},
\end{equation}
where $R_{1\pm}$ and $R_{3\pm}$ are four arbitrary functions and where $\widehat{\Omega} u_\pm = \mp(j+1/2) u_\pm$.

We now plug the results derived so far into the equations of motion in (\ref{H eom 2}).  All angular dependence cancels out, leaving behind radial equations of motion.  $u_+$ and $u_-$ each lead to four equations, one equation for each component.  In both cases, only two of the equations are independent.  Putting everything together, the radial equations of motion are
\begin{equation} 
\begin{split}
\omega R_{3\pm}
&= 
-\frac{i\alpha}{a} 
\left(\partial_r  + \frac{\partial_r \alpha}{2\alpha} + \frac{1}{r} \right) R_{1\pm}
- \alpha m_\psi R_{3\pm}
\\
&\qquad 
\pm \frac{i\alpha}{r} \left(j+\frac{1}{2}\right) R_{1\pm}
\\
\omega R_{1\pm}
&= 
-\frac{i\alpha}{a} 
\left(\partial_r  + \frac{\partial_r \alpha}{2\alpha} + \frac{1}{r} \right) R_{3\pm}
+ \alpha m_\psi R_{1\pm}
\\
&\qquad
\mp \frac{i\alpha}{r}\left(j+\frac{1}{2}\right) R_{3\pm}.
\end{split}
\end{equation}
Comparing these equations to those in (\ref{radial eom pm}), we find that they are identical.




\begin{thebibliography}{51}%
\makeatletter
\providecommand \@ifxundefined [1]{%
 \@ifx{#1\undefined}
}%
\providecommand \@ifnum [1]{%
 \ifnum #1\expandafter \@firstoftwo
 \else \expandafter \@secondoftwo
 \fi
}%
\providecommand \@ifx [1]{%
 \ifx #1\expandafter \@firstoftwo
 \else \expandafter \@secondoftwo
 \fi
}%
\providecommand \natexlab [1]{#1}%
\providecommand \enquote  [1]{``#1''}%
\providecommand \bibnamefont  [1]{#1}%
\providecommand \bibfnamefont [1]{#1}%
\providecommand \citenamefont [1]{#1}%
\providecommand \href@noop [0]{\@secondoftwo}%
\providecommand \href [0]{\begingroup \@sanitize@url \@href}%
\providecommand \@href[1]{\@@startlink{#1}\@@href}%
\providecommand \@@href[1]{\endgroup#1\@@endlink}%
\providecommand \@sanitize@url [0]{\catcode `\\12\catcode `\$12\catcode
  `\&12\catcode `\#12\catcode `\^12\catcode `\_12\catcode `\%12\relax}%
\providecommand \@@startlink[1]{}%
\providecommand \@@endlink[0]{}%
\providecommand \url  [0]{\begingroup\@sanitize@url \@url }%
\providecommand \@url [1]{\endgroup\@href {#1}{\urlprefix }}%
\providecommand \urlprefix  [0]{URL }%
\providecommand \Eprint [0]{\href }%
\providecommand \doibase [0]{https://doi.org/}%
\providecommand \selectlanguage [0]{\@gobble}%
\providecommand \bibinfo  [0]{\@secondoftwo}%
\providecommand \bibfield  [0]{\@secondoftwo}%
\providecommand \translation [1]{[#1]}%
\providecommand \BibitemOpen [0]{}%
\providecommand \bibitemStop [0]{}%
\providecommand \bibitemNoStop [0]{.\EOS\space}%
\providecommand \EOS [0]{\spacefactor3000\relax}%
\providecommand \BibitemShut  [1]{\csname bibitem#1\endcsname}%
\let\auto@bib@innerbib\@empty
\bibitem [{\citenamefont {Finster}\ \emph
  {et~al.}(1999{\natexlab{a}})\citenamefont {Finster}, \citenamefont
  {Smoller},\ and\ \citenamefont {Yau}}]{Finster:1998ws}%
  \BibitemOpen
  \bibfield  {author} {\bibinfo {author} {\bibfnamefont {F.}~\bibnamefont
  {Finster}}, \bibinfo {author} {\bibfnamefont {J.}~\bibnamefont {Smoller}},\
  and\ \bibinfo {author} {\bibfnamefont {S.-T.}\ \bibnamefont {Yau}},\
  }\bibfield  {title} {\bibinfo {title} {{Particle - like solutions of the
  Einstein-Dirac equations}},\ }\href
  {https://doi.org/10.1103/PhysRevD.59.104020} {\bibfield  {journal} {\bibinfo
  {journal} {Phys. Rev. D}\ }\textbf {\bibinfo {volume} {59}},\ \bibinfo
  {pages} {104020} (\bibinfo {year} {1999}{\natexlab{a}})},\ \Eprint
  {https://arxiv.org/abs/gr-qc/9801079} {arXiv:gr-qc/9801079 [gr-qc]}
  \BibitemShut {NoStop}%
\bibitem [{\citenamefont {Finster}\ \emph
  {et~al.}(1999{\natexlab{b}})\citenamefont {Finster}, \citenamefont
  {Smoller},\ and\ \citenamefont {Yau}}]{Finster:1998ux}%
  \BibitemOpen
  \bibfield  {author} {\bibinfo {author} {\bibfnamefont {F.}~\bibnamefont
  {Finster}}, \bibinfo {author} {\bibfnamefont {J.}~\bibnamefont {Smoller}},\
  and\ \bibinfo {author} {\bibfnamefont {S.-T.}\ \bibnamefont {Yau}},\
  }\bibfield  {title} {\bibinfo {title} {{Particle - like solutions of the
  Einstein-Dirac-Maxwell equations}},\ }\href
  {https://doi.org/10.1016/S0375-9601(99)00457-0} {\bibfield  {journal}
  {\bibinfo  {journal} {Phys. Lett. A}\ }\textbf {\bibinfo {volume} {259}},\
  \bibinfo {pages} {431} (\bibinfo {year} {1999}{\natexlab{b}})},\ \Eprint
  {https://arxiv.org/abs/gr-qc/9802012} {arXiv:gr-qc/9802012 [gr-qc]}
  \BibitemShut {NoStop}%
\bibitem [{\citenamefont {Finster}\ \emph
  {et~al.}(2000{\natexlab{a}})\citenamefont {Finster}, \citenamefont
  {Smoller},\ and\ \citenamefont {Yau}}]{Finster:1998ak}%
  \BibitemOpen
  \bibfield  {author} {\bibinfo {author} {\bibfnamefont {F.}~\bibnamefont
  {Finster}}, \bibinfo {author} {\bibfnamefont {J.}~\bibnamefont {Smoller}},\
  and\ \bibinfo {author} {\bibfnamefont {S.-T.}\ \bibnamefont {Yau}},\
  }\bibfield  {title} {\bibinfo {title} {{Nonexistence of time periodic
  solutions of the Dirac equation in a Reissner-Nordstrom black hole
  background}},\ }\href {https://doi.org/10.1063/1.533234} {\bibfield
  {journal} {\bibinfo  {journal} {J. Math. Phys.}\ }\textbf {\bibinfo {volume}
  {41}},\ \bibinfo {pages} {2173} (\bibinfo {year} {2000}{\natexlab{a}})},\
  \Eprint {https://arxiv.org/abs/gr-qc/9805050} {arXiv:gr-qc/9805050}
  \BibitemShut {NoStop}%
\bibitem [{\citenamefont {Finster}\ \emph
  {et~al.}(1999{\natexlab{c}})\citenamefont {Finster}, \citenamefont
  {Smoller},\ and\ \citenamefont {Yau}}]{Finster:1998ju}%
  \BibitemOpen
  \bibfield  {author} {\bibinfo {author} {\bibfnamefont {F.}~\bibnamefont
  {Finster}}, \bibinfo {author} {\bibfnamefont {J.}~\bibnamefont {Smoller}},\
  and\ \bibinfo {author} {\bibfnamefont {S.-T.}\ \bibnamefont {Yau}},\
  }\bibfield  {title} {\bibinfo {title} {{Nonexistence of black hole solutions
  for a spherically symmetric, static Einstein-Dirac-Maxwell system}},\ }\href
  {https://doi.org/10.1007/s002200050675} {\bibfield  {journal} {\bibinfo
  {journal} {Commun. Math. Phys.}\ }\textbf {\bibinfo {volume} {205}},\
  \bibinfo {pages} {249} (\bibinfo {year} {1999}{\natexlab{c}})},\ \Eprint
  {https://arxiv.org/abs/gr-qc/9810048} {arXiv:gr-qc/9810048} \BibitemShut
  {NoStop}%
\bibitem [{\citenamefont {Finster}\ \emph
  {et~al.}(2000{\natexlab{b}})\citenamefont {Finster}, \citenamefont
  {Smoller},\ and\ \citenamefont {Yau}}]{Finster:2000ps}%
  \BibitemOpen
  \bibfield  {author} {\bibinfo {author} {\bibfnamefont {F.}~\bibnamefont
  {Finster}}, \bibinfo {author} {\bibfnamefont {J.}~\bibnamefont {Smoller}},\
  and\ \bibinfo {author} {\bibfnamefont {S.-T.}\ \bibnamefont {Yau}},\
  }\bibfield  {title} {\bibinfo {title} {{The Interaction of Dirac particles
  with nonAbelian gauge fields and gravity bound states}},\ }\href
  {https://doi.org/10.1016/S0550-3213(00)00370-9} {\bibfield  {journal}
  {\bibinfo  {journal} {Nucl. Phys. B}\ }\textbf {\bibinfo {volume} {584}},\
  \bibinfo {pages} {387} (\bibinfo {year} {2000}{\natexlab{b}})},\ \Eprint
  {https://arxiv.org/abs/gr-qc/0001067} {arXiv:gr-qc/0001067 [gr-qc]}
  \BibitemShut {NoStop}%
\bibitem [{\citenamefont {Nodari}(2010)}]{Nodari1}%
  \BibitemOpen
  \bibfield  {author} {\bibinfo {author} {\bibfnamefont {S.~R.}\ \bibnamefont
  {Nodari}},\ }\bibfield  {title} {\bibinfo {title} {{Perturbation Method for
  Particle-like Solutions of the Einstein–Dirac Equations}},\ }\href
  {https://doi.org/10.1007/s00023-009-0015-x} {\bibfield  {journal} {\bibinfo
  {journal} {Ann. Henri Poincar\'e}\ }\textbf {\bibinfo {volume} {10}},\
  \bibinfo {pages} {1377} (\bibinfo {year} {2010})}\BibitemShut {NoStop}%
\bibitem [{\citenamefont {{Rota Nodari}}(2010)}]{Nodari2}%
  \BibitemOpen
  \bibfield  {author} {\bibinfo {author} {\bibfnamefont {S.}~\bibnamefont
  {{Rota Nodari}}},\ }\bibfield  {title} {\bibinfo {title} {Perturbation method
  for particle-like solutions of the einstein–dirac–maxwell equations},\
  }\href {https://doi.org/https://doi.org/10.1016/j.crma.2010.06.003}
  {\bibfield  {journal} {\bibinfo  {journal} {Comptes Rendus Mathematique}\
  }\textbf {\bibinfo {volume} {348}},\ \bibinfo {pages} {791} (\bibinfo {year}
  {2010})}\BibitemShut {NoStop}%
\bibitem [{\citenamefont {Stuart}(2010)}]{Stuart}%
  \BibitemOpen
  \bibfield  {author} {\bibinfo {author} {\bibfnamefont {D.}~\bibnamefont
  {Stuart}},\ }\bibfield  {title} {\bibinfo {title} {{Existence and Newtonian
  limit of nonlinear bound states in the Einstein–Dirac system}},\ }\href
  {https://doi.org/https://doi.org/10.1063/1.3294085} {\bibfield  {journal}
  {\bibinfo  {journal} {J. Math. Phys.}\ }\textbf {\bibinfo {volume} {51}},\
  \bibinfo {pages} {032501} (\bibinfo {year} {2010})}\BibitemShut {NoStop}%
\bibitem [{\citenamefont {Adanhounme}\ \emph {et~al.}(2012)\citenamefont
  {Adanhounme}, \citenamefont {Adomou}, \citenamefont {Codo},\ and\
  \citenamefont {Hounkonnou}}]{Adanhounme:2012cm}%
  \BibitemOpen
  \bibfield  {author} {\bibinfo {author} {\bibfnamefont {V.}~\bibnamefont
  {Adanhounme}}, \bibinfo {author} {\bibfnamefont {A.}~\bibnamefont {Adomou}},
  \bibinfo {author} {\bibfnamefont {F.~P.}\ \bibnamefont {Codo}},\ and\
  \bibinfo {author} {\bibfnamefont {M.~N.}\ \bibnamefont {Hounkonnou}},\
  }\bibfield  {title} {\bibinfo {title} {{Nonlinear spinor field equations in
  gravitational theory: spherical symmetric soliton-like solutions}},\ }\href
  {https://doi.org/10.4236/jmp.2012.39122} {\bibfield  {journal} {\bibinfo
  {journal} {J. Mod. Phys.}\ }\textbf {\bibinfo {volume} {3}},\ \bibinfo
  {pages} {935} (\bibinfo {year} {2012})},\ \Eprint
  {https://arxiv.org/abs/1211.3388} {arXiv:1211.3388 [math-ph]} \BibitemShut
  {NoStop}%
\bibitem [{\citenamefont {Krechet}\ and\ \citenamefont
  {Sinilshchikova}(2014)}]{Krechet:2014nda}%
  \BibitemOpen
  \bibfield  {author} {\bibinfo {author} {\bibfnamefont {V.~G.}\ \bibnamefont
  {Krechet}}\ and\ \bibinfo {author} {\bibfnamefont {I.~V.}\ \bibnamefont
  {Sinilshchikova}},\ }\bibfield  {title} {\bibinfo {title} {{Self-Gravitating
  Nonlinear Spinor Field in Stationary Spaces with Spherical Symmetry}},\
  }\href {https://doi.org/10.1007/s11182-014-0319-2} {\bibfield  {journal}
  {\bibinfo  {journal} {Russ. Phys. J.}\ }\textbf {\bibinfo {volume} {57}},\
  \bibinfo {pages} {870} (\bibinfo {year} {2014})}\BibitemShut {NoStop}%
\bibitem [{\citenamefont {Herdeiro}\ \emph {et~al.}(2017)\citenamefont
  {Herdeiro}, \citenamefont {Pombo},\ and\ \citenamefont
  {Radu}}]{Herdeiro:2017fhv}%
  \BibitemOpen
  \bibfield  {author} {\bibinfo {author} {\bibfnamefont {C.~A.~R.}\
  \bibnamefont {Herdeiro}}, \bibinfo {author} {\bibfnamefont {A.~M.}\
  \bibnamefont {Pombo}},\ and\ \bibinfo {author} {\bibfnamefont
  {E.}~\bibnamefont {Radu}},\ }\bibfield  {title} {\bibinfo {title}
  {{Asymptotically flat scalar, Dirac and Proca stars: discrete vs. continuous
  families of solutions}},\ }\href
  {https://doi.org/10.1016/j.physletb.2017.09.036} {\bibfield  {journal}
  {\bibinfo  {journal} {Phys. Lett. B}\ }\textbf {\bibinfo {volume} {773}},\
  \bibinfo {pages} {654} (\bibinfo {year} {2017})},\ \Eprint
  {https://arxiv.org/abs/1708.05674} {arXiv:1708.05674 [gr-qc]} \BibitemShut
  {NoStop}%
\bibitem [{\citenamefont {Dzhunushaliev}\ and\ \citenamefont
  {Folomeev}(2019{\natexlab{a}})}]{Dzhunushaliev:2018jhj}%
  \BibitemOpen
  \bibfield  {author} {\bibinfo {author} {\bibfnamefont {V.}~\bibnamefont
  {Dzhunushaliev}}\ and\ \bibinfo {author} {\bibfnamefont {V.}~\bibnamefont
  {Folomeev}},\ }\bibfield  {title} {\bibinfo {title} {{Dirac stars supported
  by nonlinear spinor fields}},\ }\href
  {https://doi.org/10.1103/PhysRevD.99.084030} {\bibfield  {journal} {\bibinfo
  {journal} {Phys. Rev. D}\ }\textbf {\bibinfo {volume} {99}},\ \bibinfo
  {pages} {084030} (\bibinfo {year} {2019}{\natexlab{a}})},\ \Eprint
  {https://arxiv.org/abs/1811.07500} {arXiv:1811.07500 [gr-qc]} \BibitemShut
  {NoStop}%
\bibitem [{\citenamefont {Dzhunushaliev}\ and\ \citenamefont
  {Folomeev}(2019{\natexlab{b}})}]{Dzhunushaliev:2019kiy}%
  \BibitemOpen
  \bibfield  {author} {\bibinfo {author} {\bibfnamefont {V.}~\bibnamefont
  {Dzhunushaliev}}\ and\ \bibinfo {author} {\bibfnamefont {V.}~\bibnamefont
  {Folomeev}},\ }\bibfield  {title} {\bibinfo {title} {{Dirac star in the
  presence of Maxwell and Proca fields}},\ }\href
  {https://doi.org/10.1103/PhysRevD.99.104066} {\bibfield  {journal} {\bibinfo
  {journal} {Phys. Rev. D}\ }\textbf {\bibinfo {volume} {99}},\ \bibinfo
  {pages} {104066} (\bibinfo {year} {2019}{\natexlab{b}})},\ \Eprint
  {https://arxiv.org/abs/1901.09905} {arXiv:1901.09905 [gr-qc]} \BibitemShut
  {NoStop}%
\bibitem [{\citenamefont {Bl\'azquez-Salcedo}\ \emph
  {et~al.}(2019)\citenamefont {Bl\'azquez-Salcedo}, \citenamefont {Knoll},\
  and\ \citenamefont {Radu}}]{Blazquez-Salcedo:2019qrz}%
  \BibitemOpen
  \bibfield  {author} {\bibinfo {author} {\bibfnamefont {J.~L.}\ \bibnamefont
  {Bl\'azquez-Salcedo}}, \bibinfo {author} {\bibfnamefont {C.}~\bibnamefont
  {Knoll}},\ and\ \bibinfo {author} {\bibfnamefont {E.}~\bibnamefont {Radu}},\
  }\bibfield  {title} {\bibinfo {title} {{Boson and Dirac stars in $D\geq 4$
  dimensions}},\ }\href {https://doi.org/10.1016/j.physletb.2019.04.035}
  {\bibfield  {journal} {\bibinfo  {journal} {Phys. Lett. B}\ }\textbf
  {\bibinfo {volume} {793}},\ \bibinfo {pages} {161} (\bibinfo {year}
  {2019})},\ \Eprint {https://arxiv.org/abs/1902.05851} {arXiv:1902.05851
  [gr-qc]} \BibitemShut {NoStop}%
\bibitem [{\citenamefont {Dzhunushaliev}\ \emph {et~al.}(2019)\citenamefont
  {Dzhunushaliev}, \citenamefont {Folomeev},\ and\ \citenamefont
  {Makhmudov}}]{Dzhunushaliev:2019ham}%
  \BibitemOpen
  \bibfield  {author} {\bibinfo {author} {\bibfnamefont {V.}~\bibnamefont
  {Dzhunushaliev}}, \bibinfo {author} {\bibfnamefont {V.}~\bibnamefont
  {Folomeev}},\ and\ \bibinfo {author} {\bibfnamefont {A.}~\bibnamefont
  {Makhmudov}},\ }\bibfield  {title} {\bibinfo {title} {{Non-Abelian
  Proca-Dirac-Higgs theory: Particlelike solutions and their energy
  spectrum}},\ }\href {https://doi.org/10.1103/PhysRevD.99.076009} {\bibfield
  {journal} {\bibinfo  {journal} {Phys. Rev. D}\ }\textbf {\bibinfo {volume}
  {99}},\ \bibinfo {pages} {076009} (\bibinfo {year} {2019})},\ \Eprint
  {https://arxiv.org/abs/1903.00338} {arXiv:1903.00338 [hep-th]} \BibitemShut
  {NoStop}%
\bibitem [{\citenamefont {Bronnikov}\ \emph {et~al.}(2020)\citenamefont
  {Bronnikov}, \citenamefont {Rybakov},\ and\ \citenamefont
  {Saha}}]{Bronnikov:2019nqa}%
  \BibitemOpen
  \bibfield  {author} {\bibinfo {author} {\bibfnamefont {K.~A.}\ \bibnamefont
  {Bronnikov}}, \bibinfo {author} {\bibfnamefont {Y.~P.}\ \bibnamefont
  {Rybakov}},\ and\ \bibinfo {author} {\bibfnamefont {B.}~\bibnamefont
  {Saha}},\ }\bibfield  {title} {\bibinfo {title} {{Spinor fields in spherical
  symmetry: Einstein\textendash{}Dirac and other space-times}},\ }\href
  {https://doi.org/10.1140/epjp/s13360-020-00150-z} {\bibfield  {journal}
  {\bibinfo  {journal} {Eur. Phys. J. Plus}\ }\textbf {\bibinfo {volume}
  {135}},\ \bibinfo {pages} {124} (\bibinfo {year} {2020})},\ \Eprint
  {https://arxiv.org/abs/1909.04789} {arXiv:1909.04789 [gr-qc]} \BibitemShut
  {NoStop}%
\bibitem [{\citenamefont {Bl\'azquez-Salcedo}\ and\ \citenamefont
  {Knoll}(2020)}]{Blazquez-Salcedo:2019uqq}%
  \BibitemOpen
  \bibfield  {author} {\bibinfo {author} {\bibfnamefont {J.~L.}\ \bibnamefont
  {Bl\'azquez-Salcedo}}\ and\ \bibinfo {author} {\bibfnamefont
  {C.}~\bibnamefont {Knoll}},\ }\bibfield  {title} {\bibinfo {title}
  {{Constructing spherically symmetric Einstein\textendash{}Dirac systems with
  multiple spinors: Ansatz, wormholes and other analytical solutions}},\ }\href
  {https://doi.org/10.1140/epjc/s10052-020-7706-3} {\bibfield  {journal}
  {\bibinfo  {journal} {Eur. Phys. J. C}\ }\textbf {\bibinfo {volume} {80}},\
  \bibinfo {pages} {174} (\bibinfo {year} {2020})},\ \Eprint
  {https://arxiv.org/abs/1910.03565} {arXiv:1910.03565 [gr-qc]} \BibitemShut
  {NoStop}%
\bibitem [{\citenamefont {Dzhunushaliev}\ and\ \citenamefont
  {Folomeev}(2020)}]{Dzhunushaliev:2019uft}%
  \BibitemOpen
  \bibfield  {author} {\bibinfo {author} {\bibfnamefont {V.}~\bibnamefont
  {Dzhunushaliev}}\ and\ \bibinfo {author} {\bibfnamefont {V.}~\bibnamefont
  {Folomeev}},\ }\bibfield  {title} {\bibinfo {title} {{Dirac Star with SU(2)
  Yang-Mills and Proca Fields}},\ }\href
  {https://doi.org/10.1103/PhysRevD.101.024023} {\bibfield  {journal} {\bibinfo
   {journal} {Phys. Rev. D}\ }\textbf {\bibinfo {volume} {101}},\ \bibinfo
  {pages} {024023} (\bibinfo {year} {2020})},\ \Eprint
  {https://arxiv.org/abs/1911.11614} {arXiv:1911.11614 [gr-qc]} \BibitemShut
  {NoStop}%
\bibitem [{\citenamefont {Leith}\ \emph {et~al.}(2020)\citenamefont {Leith},
  \citenamefont {Hooley}, \citenamefont {Horne},\ and\ \citenamefont
  {Dritschel}}]{Leith:2020jqw}%
  \BibitemOpen
  \bibfield  {author} {\bibinfo {author} {\bibfnamefont {P.~E.~D.}\
  \bibnamefont {Leith}}, \bibinfo {author} {\bibfnamefont {C.~A.}\ \bibnamefont
  {Hooley}}, \bibinfo {author} {\bibfnamefont {K.}~\bibnamefont {Horne}},\ and\
  \bibinfo {author} {\bibfnamefont {D.~G.}\ \bibnamefont {Dritschel}},\
  }\bibfield  {title} {\bibinfo {title} {{Fermion self-trapping in the optical
  geometry of Einstein-Dirac solitons}},\ }\href
  {https://doi.org/10.1103/PhysRevD.101.106012} {\bibfield  {journal} {\bibinfo
   {journal} {Phys. Rev. D}\ }\textbf {\bibinfo {volume} {101}},\ \bibinfo
  {pages} {106012} (\bibinfo {year} {2020})},\ \Eprint
  {https://arxiv.org/abs/2002.02747} {arXiv:2002.02747 [gr-qc]} \BibitemShut
  {NoStop}%
\bibitem [{\citenamefont {Bakucz~Can\'ario}\ \emph {et~al.}(2020)\citenamefont
  {Bakucz~Can\'ario}, \citenamefont {Lloyd}, \citenamefont {Horne},\ and\
  \citenamefont {Hooley}}]{BakuczCanario:2020qmq}%
  \BibitemOpen
  \bibfield  {author} {\bibinfo {author} {\bibfnamefont {D.}~\bibnamefont
  {Bakucz~Can\'ario}}, \bibinfo {author} {\bibfnamefont {S.}~\bibnamefont
  {Lloyd}}, \bibinfo {author} {\bibfnamefont {K.}~\bibnamefont {Horne}},\ and\
  \bibinfo {author} {\bibfnamefont {C.~A.}\ \bibnamefont {Hooley}},\ }\bibfield
   {title} {\bibinfo {title} {{Infinite-redshift localized states of Dirac
  fermions under Einsteinian gravity}},\ }\href
  {https://doi.org/10.1103/PhysRevD.102.084049} {\bibfield  {journal} {\bibinfo
   {journal} {Phys. Rev. D}\ }\textbf {\bibinfo {volume} {102}},\ \bibinfo
  {pages} {084049} (\bibinfo {year} {2020})},\ \Eprint
  {https://arxiv.org/abs/2008.10455} {arXiv:2008.10455 [gr-qc]} \BibitemShut
  {NoStop}%
\bibitem [{\citenamefont {Minamitsuji}(2020)}]{Minamitsuji:2020hpl}%
  \BibitemOpen
  \bibfield  {author} {\bibinfo {author} {\bibfnamefont {M.}~\bibnamefont
  {Minamitsuji}},\ }\bibfield  {title} {\bibinfo {title} {{Stealth spontaneous
  spinorization of relativistic stars}},\ }\href
  {https://doi.org/10.1103/PhysRevD.102.044048} {\bibfield  {journal} {\bibinfo
   {journal} {Phys. Rev. D}\ }\textbf {\bibinfo {volume} {102}},\ \bibinfo
  {pages} {044048} (\bibinfo {year} {2020})},\ \Eprint
  {https://arxiv.org/abs/2008.12758} {arXiv:2008.12758 [gr-qc]} \BibitemShut
  {NoStop}%
\bibitem [{\citenamefont {Leith}\ \emph {et~al.}(2021)\citenamefont {Leith},
  \citenamefont {Hooley}, \citenamefont {Horne},\ and\ \citenamefont
  {Dritschel}}]{Leith:2021urf}%
  \BibitemOpen
  \bibfield  {author} {\bibinfo {author} {\bibfnamefont {P.~E.~D.}\
  \bibnamefont {Leith}}, \bibinfo {author} {\bibfnamefont {C.~A.}\ \bibnamefont
  {Hooley}}, \bibinfo {author} {\bibfnamefont {K.}~\bibnamefont {Horne}},\ and\
  \bibinfo {author} {\bibfnamefont {D.~G.}\ \bibnamefont {Dritschel}},\
  }\bibfield  {title} {\bibinfo {title} {{Nonlinear effects in the excited
  states of many-fermion Einstein-Dirac solitons}},\ }\href
  {https://doi.org/10.1103/PhysRevD.104.046024} {\bibfield  {journal} {\bibinfo
   {journal} {Phys. Rev. D}\ }\textbf {\bibinfo {volume} {104}},\ \bibinfo
  {pages} {046024} (\bibinfo {year} {2021})},\ \Eprint
  {https://arxiv.org/abs/2105.12672} {arXiv:2105.12672 [gr-qc]} \BibitemShut
  {NoStop}%
\bibitem [{\citenamefont {Leith}\ \emph {et~al.}(2022)\citenamefont {Leith},
  \citenamefont {Leggat}, \citenamefont {Hooley}, \citenamefont {Horne},\ and\
  \citenamefont {Dritschel}}]{Leith:2022jew}%
  \BibitemOpen
  \bibfield  {author} {\bibinfo {author} {\bibfnamefont {P.~E.~D.}\
  \bibnamefont {Leith}}, \bibinfo {author} {\bibfnamefont {A.~D.}\ \bibnamefont
  {Leggat}}, \bibinfo {author} {\bibfnamefont {C.~A.}\ \bibnamefont {Hooley}},
  \bibinfo {author} {\bibfnamefont {K.}~\bibnamefont {Horne}},\ and\ \bibinfo
  {author} {\bibfnamefont {D.~G.}\ \bibnamefont {Dritschel}},\ }\bibfield
  {title} {\bibinfo {title} {{Gravitationally bound states of two neutral
  fermions and a Higgs field can be parametrically lighter than their
  constituents}},\ }\href@noop {} {\  (\bibinfo {year} {2022})},\ \Eprint
  {https://arxiv.org/abs/2202.03228} {arXiv:2202.03228 [gr-qc]} \BibitemShut
  {NoStop}%
\bibitem [{\citenamefont {Liang}\ \emph {et~al.}(2023)\citenamefont {Liang},
  \citenamefont {Ren}, \citenamefont {Sun},\ and\ \citenamefont
  {Wang}}]{Liang:2022mjo}%
  \BibitemOpen
  \bibfield  {author} {\bibinfo {author} {\bibfnamefont {C.}~\bibnamefont
  {Liang}}, \bibinfo {author} {\bibfnamefont {J.-R.}\ \bibnamefont {Ren}},
  \bibinfo {author} {\bibfnamefont {S.-X.}\ \bibnamefont {Sun}},\ and\ \bibinfo
  {author} {\bibfnamefont {Y.-Q.}\ \bibnamefont {Wang}},\ }\bibfield  {title}
  {\bibinfo {title} {{Dirac-boson stars}},\ }\href
  {https://doi.org/10.1007/JHEP02(2023)249} {\bibfield  {journal} {\bibinfo
  {journal} {JHEP}\ }\textbf {\bibinfo {volume} {02}},\ \bibinfo {pages}
  {249}},\ \Eprint {https://arxiv.org/abs/2207.11147} {arXiv:2207.11147
  [gr-qc]} \BibitemShut {NoStop}%
\bibitem [{\citenamefont {Bronnikov}\ \emph {et~al.}(2004)\citenamefont
  {Bronnikov}, \citenamefont {Chudaeva},\ and\ \citenamefont
  {Shikin}}]{Bronnikov:2004uu}%
  \BibitemOpen
  \bibfield  {author} {\bibinfo {author} {\bibfnamefont {K.~A.}\ \bibnamefont
  {Bronnikov}}, \bibinfo {author} {\bibfnamefont {E.~N.}\ \bibnamefont
  {Chudaeva}},\ and\ \bibinfo {author} {\bibfnamefont {G.~N.}\ \bibnamefont
  {Shikin}},\ }\bibfield  {title} {\bibinfo {title} {{Self-gravitating
  string-like configurations of nonlinear spinor fields}},\ }\href
  {https://doi.org/10.1023/B:GERG.0000032146.03936.45} {\bibfield  {journal}
  {\bibinfo  {journal} {Gen. Rel. Grav.}\ }\textbf {\bibinfo {volume} {36}},\
  \bibinfo {pages} {1537} (\bibinfo {year} {2004})}\BibitemShut {NoStop}%
\bibitem [{\citenamefont {Herdeiro}\ \emph {et~al.}(2019)\citenamefont
  {Herdeiro}, \citenamefont {Perapechka}, \citenamefont {Radu},\ and\
  \citenamefont {Shnir}}]{Herdeiro:2019mbz}%
  \BibitemOpen
  \bibfield  {author} {\bibinfo {author} {\bibfnamefont {C.}~\bibnamefont
  {Herdeiro}}, \bibinfo {author} {\bibfnamefont {I.}~\bibnamefont
  {Perapechka}}, \bibinfo {author} {\bibfnamefont {E.}~\bibnamefont {Radu}},\
  and\ \bibinfo {author} {\bibfnamefont {Y.}~\bibnamefont {Shnir}},\ }\bibfield
   {title} {\bibinfo {title} {{Asymptotically flat spinning scalar, Dirac and
  Proca stars}},\ }\href {https://doi.org/10.1016/j.physletb.2019.134845}
  {\bibfield  {journal} {\bibinfo  {journal} {Phys. Lett. B}\ }\textbf
  {\bibinfo {volume} {797}},\ \bibinfo {pages} {134845} (\bibinfo {year}
  {2019})},\ \Eprint {https://arxiv.org/abs/1906.05386} {arXiv:1906.05386
  [gr-qc]} \BibitemShut {NoStop}%
\bibitem [{\citenamefont {Bronnikov}\ and\ \citenamefont
  {Lemos}(2009)}]{Bronnikov:2009na}%
  \BibitemOpen
  \bibfield  {author} {\bibinfo {author} {\bibfnamefont {K.~A.}\ \bibnamefont
  {Bronnikov}}\ and\ \bibinfo {author} {\bibfnamefont {J.~P.~S.}\ \bibnamefont
  {Lemos}},\ }\bibfield  {title} {\bibinfo {title} {{Cylindrical wormholes}},\
  }\href {https://doi.org/10.1103/PhysRevD.79.104019} {\bibfield  {journal}
  {\bibinfo  {journal} {Phys. Rev. D}\ }\textbf {\bibinfo {volume} {79}},\
  \bibinfo {pages} {104019} (\bibinfo {year} {2009})},\ \Eprint
  {https://arxiv.org/abs/0902.2360} {arXiv:0902.2360 [gr-qc]} \BibitemShut
  {NoStop}%
\bibitem [{\citenamefont {Ventrella}\ and\ \citenamefont
  {Choptuik}(2003)}]{Ventrella:2003fu}%
  \BibitemOpen
  \bibfield  {author} {\bibinfo {author} {\bibfnamefont {J.~F.}\ \bibnamefont
  {Ventrella}}\ and\ \bibinfo {author} {\bibfnamefont {M.~W.}\ \bibnamefont
  {Choptuik}},\ }\bibfield  {title} {\bibinfo {title} {{Critical phenomena in
  the Einstein massless Dirac system}},\ }\href
  {https://doi.org/10.1103/PhysRevD.68.044020} {\bibfield  {journal} {\bibinfo
  {journal} {Phys. Rev. D}\ }\textbf {\bibinfo {volume} {68}},\ \bibinfo
  {pages} {044020} (\bibinfo {year} {2003})},\ \Eprint
  {https://arxiv.org/abs/gr-qc/0304007} {arXiv:gr-qc/0304007 [gr-qc]}
  \BibitemShut {NoStop}%
\bibitem [{\citenamefont {Zeller}\ and\ \citenamefont
  {Hiptmair}(2006)}]{Zeller:2006rm}%
  \BibitemOpen
  \bibfield  {author} {\bibinfo {author} {\bibfnamefont {B.}~\bibnamefont
  {Zeller}}\ and\ \bibinfo {author} {\bibfnamefont {R.}~\bibnamefont
  {Hiptmair}},\ }\bibfield  {title} {\bibinfo {title} {{Conservative
  discretization of the Einstein-Dirac equations in spherically symmetric
  spacetime}},\ }\href {https://doi.org/10.1088/0264-9381/23/16/S17} {\bibfield
   {journal} {\bibinfo  {journal} {Class. Quant. Grav.}\ }\textbf {\bibinfo
  {volume} {23}},\ \bibinfo {pages} {S615} (\bibinfo {year} {2006})},\ \Eprint
  {https://arxiv.org/abs/gr-qc/0607128} {arXiv:gr-qc/0607128 [gr-qc]}
  \BibitemShut {NoStop}%
\bibitem [{\citenamefont {Daka}\ \emph {et~al.}(2019)\citenamefont {Daka},
  \citenamefont {Phan},\ and\ \citenamefont {Kain}}]{Daka:2019iix}%
  \BibitemOpen
  \bibfield  {author} {\bibinfo {author} {\bibfnamefont {E.}~\bibnamefont
  {Daka}}, \bibinfo {author} {\bibfnamefont {N.~N.}\ \bibnamefont {Phan}},\
  and\ \bibinfo {author} {\bibfnamefont {B.}~\bibnamefont {Kain}},\ }\bibfield
  {title} {\bibinfo {title} {{Perturbing the ground state of Dirac stars}},\
  }\href {https://doi.org/10.1103/PhysRevD.100.084042} {\bibfield  {journal}
  {\bibinfo  {journal} {Phys. Rev. D}\ }\textbf {\bibinfo {volume} {100}},\
  \bibinfo {pages} {084042} (\bibinfo {year} {2019})},\ \Eprint
  {https://arxiv.org/abs/1910.09415} {arXiv:1910.09415 [gr-qc]} \BibitemShut
  {NoStop}%
\bibitem [{\citenamefont {Bl\'azquez-Salcedo}\ \emph
  {et~al.}(2021)\citenamefont {Bl\'azquez-Salcedo}, \citenamefont {Knoll},\
  and\ \citenamefont {Radu}}]{Blazquez-Salcedo:2020czn}%
  \BibitemOpen
  \bibfield  {author} {\bibinfo {author} {\bibfnamefont {J.~L.}\ \bibnamefont
  {Bl\'azquez-Salcedo}}, \bibinfo {author} {\bibfnamefont {C.}~\bibnamefont
  {Knoll}},\ and\ \bibinfo {author} {\bibfnamefont {E.}~\bibnamefont {Radu}},\
  }\bibfield  {title} {\bibinfo {title} {{Traversable wormholes in
  Einstein-Dirac-Maxwell theory}},\ }\href
  {https://doi.org/10.1103/PhysRevLett.126.101102} {\bibfield  {journal}
  {\bibinfo  {journal} {Phys. Rev. Lett.}\ }\textbf {\bibinfo {volume} {126}},\
  \bibinfo {pages} {101102} (\bibinfo {year} {2021})},\ \Eprint
  {https://arxiv.org/abs/2010.07317} {arXiv:2010.07317 [gr-qc]} \BibitemShut
  {NoStop}%
\bibitem [{\citenamefont {Konoplya}\ and\ \citenamefont
  {Zhidenko}(2022)}]{Konoplya:2021hsm}%
  \BibitemOpen
  \bibfield  {author} {\bibinfo {author} {\bibfnamefont {R.~A.}\ \bibnamefont
  {Konoplya}}\ and\ \bibinfo {author} {\bibfnamefont {A.}~\bibnamefont
  {Zhidenko}},\ }\bibfield  {title} {\bibinfo {title} {{Traversable Wormholes
  in General Relativity}},\ }\href
  {https://doi.org/10.1103/PhysRevLett.128.091104} {\bibfield  {journal}
  {\bibinfo  {journal} {Phys. Rev. Lett.}\ }\textbf {\bibinfo {volume} {128}},\
  \bibinfo {pages} {091104} (\bibinfo {year} {2022})},\ \Eprint
  {https://arxiv.org/abs/2106.05034} {arXiv:2106.05034 [gr-qc]} \BibitemShut
  {NoStop}%
\bibitem [{\citenamefont {Bl\'azquez-Salcedo}\ \emph
  {et~al.}(2022)\citenamefont {Bl\'azquez-Salcedo}, \citenamefont {Knoll},\
  and\ \citenamefont {Radu}}]{Blazquez-Salcedo:2021udn}%
  \BibitemOpen
  \bibfield  {author} {\bibinfo {author} {\bibfnamefont {J.~L.}\ \bibnamefont
  {Bl\'azquez-Salcedo}}, \bibinfo {author} {\bibfnamefont {C.}~\bibnamefont
  {Knoll}},\ and\ \bibinfo {author} {\bibfnamefont {E.}~\bibnamefont {Radu}},\
  }\bibfield  {title} {\bibinfo {title}
  {{Einstein\textendash{}Dirac\textendash{}Maxwell wormholes: ansatz,
  construction and properties of symmetric solutions}},\ }\href
  {https://doi.org/10.1140/epjc/s10052-022-10488-6} {\bibfield  {journal}
  {\bibinfo  {journal} {Eur. Phys. J. C}\ }\textbf {\bibinfo {volume} {82}},\
  \bibinfo {pages} {533} (\bibinfo {year} {2022})},\ \Eprint
  {https://arxiv.org/abs/2108.12187} {arXiv:2108.12187 [gr-qc]} \BibitemShut
  {NoStop}%
\bibitem [{\citenamefont {Birrell}\ and\ \citenamefont
  {Davies}(1984)}]{Birrell:1982ix}%
  \BibitemOpen
  \bibfield  {author} {\bibinfo {author} {\bibfnamefont {N.~D.}\ \bibnamefont
  {Birrell}}\ and\ \bibinfo {author} {\bibfnamefont {P.~C.~W.}\ \bibnamefont
  {Davies}},\ }\href@noop {} {\emph {\bibinfo {title} {{Quantum Fields in
  Curved Space}}}}\ (\bibinfo  {publisher} {Cambridge Univ. Press},\ \bibinfo
  {year} {1984})\BibitemShut {NoStop}%
\bibitem [{\citenamefont {Wald}(1995)}]{Wald:1995yp}%
  \BibitemOpen
  \bibfield  {author} {\bibinfo {author} {\bibfnamefont {R.~M.}\ \bibnamefont
  {Wald}},\ }\href@noop {} {\emph {\bibinfo {title} {{Quantum Field Theory in
  Curved Space-Time and Black Hole Thermodynamics}}}},\ Chicago Lectures in
  Physics\ (\bibinfo  {publisher} {University of Chicago Press},\ \bibinfo
  {year} {1995})\BibitemShut {NoStop}%
\bibitem [{\citenamefont {Parker}\ and\ \citenamefont
  {Toms}(2009)}]{Parker:2009uva}%
  \BibitemOpen
  \bibfield  {author} {\bibinfo {author} {\bibfnamefont {L.~E.}\ \bibnamefont
  {Parker}}\ and\ \bibinfo {author} {\bibfnamefont {D.}~\bibnamefont {Toms}},\
  }\href {https://doi.org/10.1017/CBO9780511813924} {\emph {\bibinfo {title}
  {{Quantum Field Theory in Curved Spacetime}: {Quantized Field and
  Gravity}}}},\ Cambridge Monographs on Mathematical Physics\ (\bibinfo
  {publisher} {Cambridge University Press},\ \bibinfo {year}
  {2009})\BibitemShut {NoStop}%
\bibitem [{\citenamefont {Hu}\ and\ \citenamefont
  {Verdaguer}(2020)}]{Hu:2020luk}%
  \BibitemOpen
  \bibfield  {author} {\bibinfo {author} {\bibfnamefont {B.-L.~B.}\
  \bibnamefont {Hu}}\ and\ \bibinfo {author} {\bibfnamefont {E.}~\bibnamefont
  {Verdaguer}},\ }\href {https://doi.org/10.1017/9780511667497} {\emph
  {\bibinfo {title} {{Semiclassical and Stochastic Gravity}: {Quantum Field
  Effects on Curved Spacetime}}}},\ Cambridge Monographs on Mathematical
  Physics\ (\bibinfo  {publisher} {Cambridge University Press},\ \bibinfo
  {year} {2020})\BibitemShut {NoStop}%
\bibitem [{\citenamefont {Alcubierre}\ \emph {et~al.}(2023)\citenamefont
  {Alcubierre}, \citenamefont {Barranco}, \citenamefont {Bernal}, \citenamefont
  {Degollado}, \citenamefont {Diez-Tejedor}, \citenamefont {Megevand},
  \citenamefont {N\'u\~nez},\ and\ \citenamefont
  {Sarbach}}]{Alcubierre:2022rgp}%
  \BibitemOpen
  \bibfield  {author} {\bibinfo {author} {\bibfnamefont {M.}~\bibnamefont
  {Alcubierre}}, \bibinfo {author} {\bibfnamefont {J.}~\bibnamefont
  {Barranco}}, \bibinfo {author} {\bibfnamefont {A.}~\bibnamefont {Bernal}},
  \bibinfo {author} {\bibfnamefont {J.~C.}\ \bibnamefont {Degollado}}, \bibinfo
  {author} {\bibfnamefont {A.}~\bibnamefont {Diez-Tejedor}}, \bibinfo {author}
  {\bibfnamefont {M.}~\bibnamefont {Megevand}}, \bibinfo {author}
  {\bibfnamefont {D.}~\bibnamefont {N\'u\~nez}},\ and\ \bibinfo {author}
  {\bibfnamefont {O.}~\bibnamefont {Sarbach}},\ }\bibfield  {title} {\bibinfo
  {title} {{Boson stars and their relatives in semiclassical gravity}},\ }\href
  {https://doi.org/10.1103/PhysRevD.107.045017} {\bibfield  {journal} {\bibinfo
   {journal} {Phys. Rev. D}\ }\textbf {\bibinfo {volume} {107}},\ \bibinfo
  {pages} {045017} (\bibinfo {year} {2023})},\ \Eprint
  {https://arxiv.org/abs/2212.02530} {arXiv:2212.02530 [gr-qc]} \BibitemShut
  {NoStop}%
\bibitem [{\citenamefont {Kaup}(1968)}]{Kaup:1968zz}%
  \BibitemOpen
  \bibfield  {author} {\bibinfo {author} {\bibfnamefont {D.~J.}\ \bibnamefont
  {Kaup}},\ }\bibfield  {title} {\bibinfo {title} {{Klein-Gordon Geon}},\
  }\href {https://doi.org/10.1103/PhysRev.172.1331} {\bibfield  {journal}
  {\bibinfo  {journal} {Phys. Rev.}\ }\textbf {\bibinfo {volume} {172}},\
  \bibinfo {pages} {1331} (\bibinfo {year} {1968})}\BibitemShut {NoStop}%
\bibitem [{\citenamefont {Ruffini}\ and\ \citenamefont
  {Bonazzola}(1969)}]{Ruffini:1969qy}%
  \BibitemOpen
  \bibfield  {author} {\bibinfo {author} {\bibfnamefont {R.}~\bibnamefont
  {Ruffini}}\ and\ \bibinfo {author} {\bibfnamefont {S.}~\bibnamefont
  {Bonazzola}},\ }\bibfield  {title} {\bibinfo {title} {{Systems of
  selfgravitating particles in general relativity and the concept of an
  equation of state}},\ }\href {https://doi.org/10.1103/PhysRev.187.1767}
  {\bibfield  {journal} {\bibinfo  {journal} {Phys. Rev.}\ }\textbf {\bibinfo
  {volume} {187}},\ \bibinfo {pages} {1767} (\bibinfo {year}
  {1969})}\BibitemShut {NoStop}%
\bibitem [{\citenamefont {Liebling}\ and\ \citenamefont
  {Palenzuela}(2012)}]{Liebling:2012fv}%
  \BibitemOpen
  \bibfield  {author} {\bibinfo {author} {\bibfnamefont {S.~L.}\ \bibnamefont
  {Liebling}}\ and\ \bibinfo {author} {\bibfnamefont {C.}~\bibnamefont
  {Palenzuela}},\ }\bibfield  {title} {\bibinfo {title} {{Dynamical Boson
  Stars}},\ }\href {https://doi.org/10.12942/lrr-2012-6} {\bibfield  {journal}
  {\bibinfo  {journal} {Living Rev. Rel.}\ }\textbf {\bibinfo {volume} {15}},\
  \bibinfo {pages} {6} (\bibinfo {year} {2012})},\ \Eprint
  {https://arxiv.org/abs/1202.5809} {arXiv:1202.5809 [gr-qc]} \BibitemShut
  {NoStop}%
\bibitem [{\citenamefont {Alcubierre}\ \emph {et~al.}(2018)\citenamefont
  {Alcubierre}, \citenamefont {Barranco}, \citenamefont {Bernal}, \citenamefont
  {Degollado}, \citenamefont {Diez-Tejedor}, \citenamefont {Megevand},
  \citenamefont {Nunez},\ and\ \citenamefont {Sarbach}}]{Alcubierre:2018ahf}%
  \BibitemOpen
  \bibfield  {author} {\bibinfo {author} {\bibfnamefont {M.}~\bibnamefont
  {Alcubierre}}, \bibinfo {author} {\bibfnamefont {J.}~\bibnamefont
  {Barranco}}, \bibinfo {author} {\bibfnamefont {A.}~\bibnamefont {Bernal}},
  \bibinfo {author} {\bibfnamefont {J.~C.}\ \bibnamefont {Degollado}}, \bibinfo
  {author} {\bibfnamefont {A.}~\bibnamefont {Diez-Tejedor}}, \bibinfo {author}
  {\bibfnamefont {M.}~\bibnamefont {Megevand}}, \bibinfo {author}
  {\bibfnamefont {D.}~\bibnamefont {Nunez}},\ and\ \bibinfo {author}
  {\bibfnamefont {O.}~\bibnamefont {Sarbach}},\ }\bibfield  {title} {\bibinfo
  {title} {{$\ell$-Boson stars}},\ }\href
  {https://doi.org/10.1088/1361-6382/aadcb6} {\bibfield  {journal} {\bibinfo
  {journal} {Class. Quant. Grav.}\ }\textbf {\bibinfo {volume} {35}},\ \bibinfo
  {pages} {19LT01} (\bibinfo {year} {2018})},\ \Eprint
  {https://arxiv.org/abs/1805.11488} {arXiv:1805.11488 [gr-qc]} \BibitemShut
  {NoStop}%
\bibitem [{\citenamefont {Alcubierre}(2008)}]{AlcubierreBook}%
  \BibitemOpen
  \bibfield  {author} {\bibinfo {author} {\bibfnamefont {M.}~\bibnamefont
  {Alcubierre}},\ }\href@noop {} {\emph {\bibinfo {title} {{Introduction to 3+1
  numerical relativity}}}}\ (\bibinfo  {publisher} {Oxford},\ \bibinfo {year}
  {2008})\BibitemShut {NoStop}%
\bibitem [{\citenamefont {Weinberg}(1972)}]{Weinberg:1972kfs}%
  \BibitemOpen
  \bibfield  {author} {\bibinfo {author} {\bibfnamefont {S.}~\bibnamefont
  {Weinberg}},\ }\href@noop {} {\emph {\bibinfo {title} {{Gravitation and
  Cosmology}}}}\ (\bibinfo  {publisher} {John Wiley and Sons},\ \bibinfo {year}
  {1972})\BibitemShut {NoStop}%
\bibitem [{\citenamefont {Carroll}(2004)}]{Carroll:2004st}%
  \BibitemOpen
  \bibfield  {author} {\bibinfo {author} {\bibfnamefont {S.~M.}\ \bibnamefont
  {Carroll}},\ }\href@noop {} {\emph {\bibinfo {title} {{Spacetime and
  geometry: An introduction to general relativity}}}}\ (\bibinfo  {publisher}
  {Addison-Wesley},\ \bibinfo {year} {2004})\BibitemShut {NoStop}%
\bibitem [{\citenamefont {Sakurai}(1967)}]{Sakurai}%
  \BibitemOpen
  \bibfield  {author} {\bibinfo {author} {\bibfnamefont {J.~J.}\ \bibnamefont
  {Sakurai}},\ }\href@noop {} {\emph {\bibinfo {title} {{Advanced Quantum
  Mechanics}}}}\ (\bibinfo  {publisher} {Addison-Wesley},\ \bibinfo {year}
  {1967})\BibitemShut {NoStop}%
\bibitem [{\citenamefont {Bransden}\ and\ \citenamefont
  {Joachain}(2000)}]{Bransden}%
  \BibitemOpen
  \bibfield  {author} {\bibinfo {author} {\bibfnamefont {B.~H.}\ \bibnamefont
  {Bransden}}\ and\ \bibinfo {author} {\bibfnamefont {C.~J.}\ \bibnamefont
  {Joachain}},\ }\href@noop {} {\emph {\bibinfo {title} {{Quantum
  Mechanics}}}},\ \bibinfo {edition} {2nd}\ ed.\ (\bibinfo  {publisher}
  {Prentice Hall},\ \bibinfo {year} {2000})\BibitemShut {NoStop}%
\bibitem [{\citenamefont {Olabarrieta}\ \emph {et~al.}(2007)\citenamefont
  {Olabarrieta}, \citenamefont {Ventrella}, \citenamefont {Choptuik},\ and\
  \citenamefont {Unruh}}]{Olabarrieta:2007di}%
  \BibitemOpen
  \bibfield  {author} {\bibinfo {author} {\bibfnamefont {I.}~\bibnamefont
  {Olabarrieta}}, \bibinfo {author} {\bibfnamefont {J.~F.}\ \bibnamefont
  {Ventrella}}, \bibinfo {author} {\bibfnamefont {M.~W.}\ \bibnamefont
  {Choptuik}},\ and\ \bibinfo {author} {\bibfnamefont {W.~G.}\ \bibnamefont
  {Unruh}},\ }\bibfield  {title} {\bibinfo {title} {{Critical Behavior in the
  Gravitational Collapse of a Scalar Field with Angular Momentum in Spherical
  Symmetry}},\ }\href {https://doi.org/10.1103/PhysRevD.76.124014} {\bibfield
  {journal} {\bibinfo  {journal} {Phys. Rev. D}\ }\textbf {\bibinfo {volume}
  {76}},\ \bibinfo {pages} {124014} (\bibinfo {year} {2007})},\ \Eprint
  {https://arxiv.org/abs/0708.0513} {arXiv:0708.0513 [gr-qc]} \BibitemShut
  {NoStop}%
\bibitem [{\citenamefont {Diez-Tejedor}\ and\ \citenamefont
  {Sudarsky}(2012)}]{Diez-Tejedor:2011plw}%
  \BibitemOpen
  \bibfield  {author} {\bibinfo {author} {\bibfnamefont {A.}~\bibnamefont
  {Diez-Tejedor}}\ and\ \bibinfo {author} {\bibfnamefont {D.}~\bibnamefont
  {Sudarsky}},\ }\bibfield  {title} {\bibinfo {title} {{Towards a formal
  description of the collapse approach to the inflationary origin of the seeds
  of cosmic structure}},\ }\href
  {https://doi.org/10.1088/1475-7516/2012/07/045} {\bibfield  {journal}
  {\bibinfo  {journal} {JCAP}\ }\textbf {\bibinfo {volume} {07}},\ \bibinfo
  {pages} {045}},\ \Eprint {https://arxiv.org/abs/1108.4928} {arXiv:1108.4928
  [gr-qc]} \BibitemShut {NoStop}%
\bibitem [{\citenamefont {Newman}\ and\ \citenamefont
  {Penrose}(1966)}]{Newman:1966ub}%
  \BibitemOpen
  \bibfield  {author} {\bibinfo {author} {\bibfnamefont {E.~T.}\ \bibnamefont
  {Newman}}\ and\ \bibinfo {author} {\bibfnamefont {R.}~\bibnamefont
  {Penrose}},\ }\bibfield  {title} {\bibinfo {title} {{Note on the
  Bondi-Metzner-Sachs group}},\ }\href {https://doi.org/10.1063/1.1931221}
  {\bibfield  {journal} {\bibinfo  {journal} {J. Math. Phys.}\ }\textbf
  {\bibinfo {volume} {7}},\ \bibinfo {pages} {863} (\bibinfo {year}
  {1966})}\BibitemShut {NoStop}%
\bibitem [{\citenamefont {Penrose}\ and\ \citenamefont
  {Rindler}(1988)}]{Penrose:1986ca}%
  \BibitemOpen
  \bibfield  {author} {\bibinfo {author} {\bibfnamefont {R.}~\bibnamefont
  {Penrose}}\ and\ \bibinfo {author} {\bibfnamefont {W.}~\bibnamefont
  {Rindler}},\ }\href@noop {} {\emph {\bibinfo {title} {{Spinors and
  space-time}}}}\ (\bibinfo  {publisher} {Cambridge University Press},\
  \bibinfo {year} {1988})\BibitemShut {NoStop}%
\end{thebibliography}

%

\end{document}